\documentclass[prl,amsmath,amssymb,notitlepage,twocolumn,nofootinbib,superscriptaddress,longbibliography]{revtex4-2}
\usepackage{amsmath,amssymb,wasysym,graphicx}
\usepackage{times}
\usepackage[varg]{txfonts}
\usepackage{textcomp}
\usepackage{subfigure}
\usepackage{tabu}
\usepackage{color}
\usepackage{xcolor}
\usepackage[colorlinks=true,citecolor=blue,urlcolor=blue,linkcolor=blue,hyperindex]{hyperref}
\usepackage{braket}
\usepackage{dsfont}
\usepackage{overpic}
\usepackage{clrscode3e}

\usepackage[normalem]{ulem}
\usepackage{verbatim}
\allowdisplaybreaks

\newcommand{\newsect}[1]{\noindent \textit{\textcolor{blue}{#1.--}}}

\begin{document}

\title{Unraveling the Kagome Antiferromagnetic $3J$ Model and Its Materials: An Integrated Approach}

\author{Xin Lu}
\affiliation{Department of Physics and HK Institute of Quantum Science \& Technology, The University of Hong Kong, Pokfulam Road,  Hong Kong SAR, China}
\affiliation{State Key Laboratory of Optical Quantum Materials, The University of Hong Kong, Pokfulam Road,  Hong Kong SAR, China}

\author{Andreas Raikos}
\affiliation{Univ Toulouse, CNRS, Laboratoire de Physique Th\'eorique, Toulouse,  France}

\author{Menghan Song}
\affiliation{Department of Physics and HK Institute of Quantum Science \& Technology, The University of Hong Kong, Pokfulam Road,  Hong Kong SAR, China}
\affiliation{State Key Laboratory of Optical Quantum Materials, The University of Hong Kong, Pokfulam Road,  Hong Kong SAR, China}

\author{Zezong Li}
\affiliation{Beijing National Laboratory for Condensed Matter Physics, Institute of Physics, Chinese Academy of Sciences, Beijing 100190, China}

\author{Lankun Han}
\affiliation{Beijing National Laboratory for Condensed Matter Physics, Institute of Physics, Chinese Academy of Sciences, Beijing 100190, China}
\affiliation{School of Physical Sciences, University of Chinese Academy of Sciences, Beijing 100190, China}

\author{Shiliang Li}
\affiliation{Beijing National Laboratory for Condensed Matter Physics, Institute of Physics, Chinese Academy of Sciences, Beijing 100190, China}
\affiliation{School of Physical Sciences, University of Chinese Academy of Sciences, Beijing 100190, China}

\author{Sylvain Capponi}
\affiliation{Univ Toulouse, CNRS, Laboratoire de Physique Th\'eorique, Toulouse,  France}

\author{Zi Yang Meng}
\email{zymeng@hku.hk}
\affiliation{Department of Physics and HK Institute of Quantum Science \& Technology, The University of Hong Kong, Pokfulam Road,  Hong Kong SAR, China}
\affiliation{State Key Laboratory of Optical Quantum Materials, The University of Hong Kong, Pokfulam Road,  Hong Kong SAR, China}

\author{Chengkang Zhou}
\email{zhouchk2@hku.hk}
\affiliation{Department of Physics and HK Institute of Quantum Science \& Technology, The University of Hong Kong, Pokfulam Road,  Hong Kong SAR, China}
\affiliation{State Key Laboratory of Optical Quantum Materials, The University of Hong Kong, Pokfulam Road,  Hong Kong SAR, China}

\date{\today}
\begin{abstract}
We investigate the ground-state and finite-temperature properties of the kagome antiferromagnetic Heisenberg model with three inequivalent couplings, 
dubbed the $3J$ model, which is designed for the candidate Dirac quantum spin liquid (QSL) material YCu$_3$(OH)$_6$Br$_2$[Br$_{1-x}$(OH)$_x$]~\cite{Spectralevidence2024Zeng}. Employing large-scale density-matrix renormalization group (DMRG) supplemented by neural quantum states (NQS) simulations, we identify an intermediate QSL phase between two magnetically ordered phases.
We also find that this QSL is separated from the kagome spin liquid ground state at the isotropic limit. To establish a direct comparison with experiments, we compute the specific heat of the model by means of advanced exponential (XTRG) and tangent-space (tanTRG) thermal tensor-network methods. In the magnetically ordered phase, the specific heat over temperature exhibits a shoulder at a temperature that is a fraction of the coupling strength $J_{\hexagon}$, which disappears in the QSL phase. These universal behaviors are consistent with the experimentally observed specific heat in $3J$ materials for both ordered and QSL candidate samples. Our work thus connects microscopic models with experimentally measurable signatures, exemplifying an integrated approach to understanding QSL phenomena in frustrated quantum magnets~\cite{mengPerspective26}, with $3J$ materials serving as a representative case and providing a foundation for future studies.
%Our work thus connects microscopic models with experimentally measurable signatures and offers a representative example of the integrated approach for studying frustrated quantum magnets~\cite{mengPerspective26}.
\end{abstract}
\maketitle

\newsect{Introduction} Quantum magnets provide a fertile ground for exploring exotic quantum phases, among which quantum spin liquids (QSLs)~\cite{SavaryL17,BalentsL10,knolleField2019,BroholmC20,mengPerspective26,Montecarlo2019Xu} are particularly intriguing and elusive due to their departure from conventional ordered states. Rather than developing conventional long-range order, QSLs can host long-range quantum entanglement, fractionalized excitations, and emergent gauge structures, providing a natural setting for quantum phases beyond the Landau paradigm of symmetry breaking. One promising platform for realizing QSLs is the highly frustrated spin-$1/2$ kagome antiferromagnetic Heisenberg model~\cite{HastingsMB00,HermeleM08,ZhuW18,ZhuW19,IqbalY21,KieseD23,FerrariF24,ZWGSSDNS_2025QF}. Despite extensive numerical and theoretical efforts, however, the nature of its ground state remains debated, with various studies supporting competing scenarios, e.g., a gapped QSL~\cite{JHC_PRL2008,YanS11,DepenbrockS12,JHC_NP2012,Messio_PRL2012,GSS_SR2014,li2016z2spinliquidphase,JWM_PRB2017,SRY_npjQM2024}, a gapless QSL~\cite{RanY07,Iqbal_PRB2011,IqbalBecca_PRB2013,IqbalY14,Hu_PRB2015,HeYC17,LHJXT_PRL2017,JiangSH_2019SciPostPhys} or valence-bond
crystal (VBC) states~\cite{Budnik_PRL2004,SinghVBC_PRB2007,Capponi2013,Evenbly_PRL2020}. This theoretical ambiguity is compounded by experimental challenges: disorder and magnetic impurities in candidate kagome materials~\cite{VriesMA08,FreedmanDE10,YYHuang2021,HanTH12,FuM15,HanTH16,KhuntiaP20,FengZL17,WeiYuan2017,FengZL18b,Dynamicfingerprint2021Fu,WeiY21,NormanMR16} can mask intrinsic signatures of QSLs in thermodynamic and spectroscopic measurements and, in some cases, may destabilize the underlying quantum phase.

%On the other hand, identifying QSLs in real materials calls for an integrated approach~\cite{mengPerspective26}, in which theoretical, numerical, and experimental work in concert to constrain and advance our understanding of the underlying quantum states. This synergy is particularly important for frustrated magnets, where finite-size and geometry limitations of numerical methods, uncertainties in microscopic model parameters, and the lack of definitive experimental ``smoking-gun'' signatures can obscure the nature of the ground state. An integrated investigation addresses these challenges through complementary probes: microscopic calculations establish the relevant model and its quantum phase structure, unbiased many-body simulations connect the model to measurable observables, while experiments provide independent tests of the theoretical description. Such a bidirectional interplay not only strengthens the identification of candidate QSL phases but also enables quantitative connections between microscopic physics and experimentally accessible phenomena. It is such a research scheme that we will follow in this work.

\begin{figure}[htp!]
\includegraphics[width=\columnwidth,angle=0]{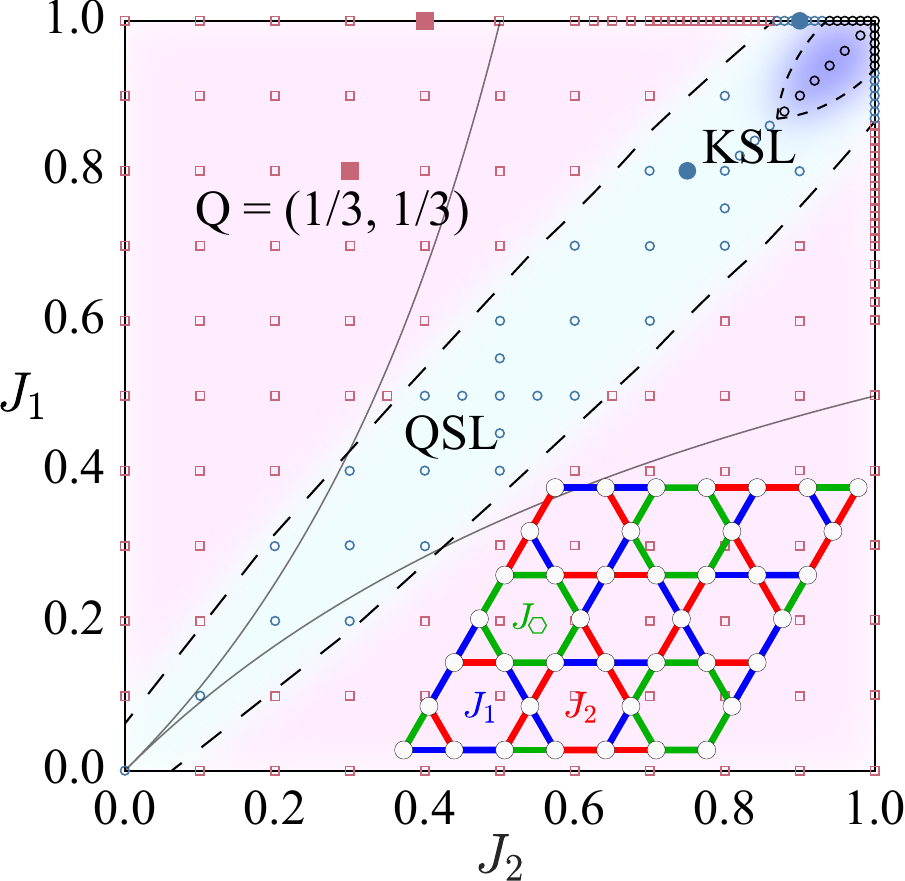}
\caption{\textbf{Ground-state phase diagram of the 3$J$ model.} Two $(1/3,1/3)$ ordered phases and an intermediate QSL phase are identified within $0\le J_{1,2}\le1$. The QSL phase is separated from the KSL ground state around the isotropic limit, $J_{1}=J_{2}=J_{\hexagon}$. Open symbols denote the calculated parameter points, while solid symbols indicate the parameter points used for specific-heat calculations. The black dashed lines indicate the quantum phase boundaries determined in this work, and the gray solid lines show the classical phase boundaries obtained from Ref.~\cite{HeringM22} for comparison. The inset displays an enlarged 27-site unit cell (in the $(1/3,1/3)$ ordered phase) of the $3J$ model on the kagome lattice, with different bond colors indicating the distinct Heisenberg interactions defined in Eq.~(\ref{eq:model}).}
\label{PDDMRG}
\end{figure}

Recent experimental advances in kagome antiferromagnetic compounds YCu$_3$(OH)$_{6+x}$X$_{3-x}$ (X=Cl or Br)~\cite{Li2025Recent, Spectralevidence2024Zeng, Possibledirac2022Zeng, liuGapless2022, XuA24, Spin2025Han, Unconventionalmagnetic2025Zheng, Antiferromagnetic2025Zezhong, Magneticordering2021Sun,ChatterjeeD23,abbruciati2026ab} have opened a promising avenue toward realizing a Dirac QSL with gapless Dirac excitations~\cite{SavaryL17,BalentsL10}. Rather than forming an isotropic kagome lattice, these materials exhibit a weakly distorted kagome geometry with three inequivalent nearest-neighbor couplings, $J_1$, $J_2$, and $J_{\hexagon}$, defining the so-called $3J$ model, as illustrated in the inset of Fig.~\ref{PDDMRG}. This distortion originates from the displacement of Y$^{3+}$ ions out of the kagome plane by the polar (OH)$^-$ ligand, which splits the uniform hexagon coupling into two alternating couplings, $J_1$ and $J_2$~\cite{liuGapless2022,XuA24}. Experiments on candidate $3J$ materials have revealed the $(1/3,1/3)$ ordered phase~\cite{Magneticordering2021Sun,ChatterjeeD23} and provided evidence for a possible Dirac QSL from specific-heat~\cite{Possibledirac2022Zeng,Antiferromagnetic2025Zezhong,Magneticordering2021Sun} and inelastic neutron scattering~\cite{Spectralevidence2024Zeng,Spin2025Han} measurements. Despite these experimental indications, theoretical understanding of the $3J$ model remains at an early stage. While the classical phase diagram has been established~\cite{HeringM22} and recent simulations have provided only a preliminary characterization of the excitation spectrum, with agreement with experiment largely limited to the high-energy regime~\cite{Spin2025Han}. More fundamentally, unbiased numerical simulations of the phase diagram of the $3J$ model, particularly the nature and stability of the putative QSL, is still largely absent. This leaves several essential questions unresolved: Does the QSL survive in the presence of the bond anisotropy intrinsic to the $3J$ model? How does it relate to the kagome spin liquid (KSL) of the isotropic ($1J$) model? Do they represent the same quantum phase, or does the bond anisotropy stabilize a distinct QSL? How does it evolve into the neighboring $(1/3,1/3)$ ordered phases? Moreover, a systematic understanding of its finite-temperature properties and their connection to experimental observations remains elusive. Addressing these questions is essential for establishing whether the $3J$ model can provide a robust theoretical framework for the experimentally observed QSL phenomena.

Motivated by the integrated strategy~\cite{mengPerspective26}, in this Letter we combine four state-of-the-art quantum many-body computational approaches with measurements on multiple experimental samples to investigate the kagome antiferromagnetic $3J$ model, providing a concrete demonstration of the strength of such an integrated strategy. Rather than relying on any single numerical method or experimental signature, we establish the ground-state phase diagram via complementary large-scale density-matrix renormalization group (DMRG)~\cite{dmrg_white_1992,dmrg_white_1993,dmrg_ulrich_2005,dmrg_ulrich_2011} and neural quantum states (NQS)~\cite{Carleo2017,Lange_2024, Rigo_2026,Viteritti2025} calculations, characterize the finite-temperature behavior using advanced exponential (XTRG)~\cite{Chen2018Exponential} and tangent-space (tanTRG)~\cite{Liu2023Tangent} thermal tensor-network methods, and compare the resulting numerical predictions against specific-heat measurements from multiple samples, represented by the ordered Y$_3$Cu$_9$(OH)$_{19}$Cl$_8$~\cite{Magneticordering2021Sun} and QSL candidate materials  YCu$_3$(OH)$_6$Br$_2$[Br$_{1-x}$(OH)$_x$]~\cite{Spectralevidence2024Zeng,Possibledirac2022Zeng,XuA24} and LuCu$_3$(OH)$_6$Br$_2$[Br$_x$(OH)$_{1-x}$]~\cite{Antiferromagnetic2025Zezhong} samples.

This combined strategy not only allows us to exploit the respective strengths of different numerical approaches using experiment as an independent touchstone, but also enables us to cross-validate and synthesize their insights to uncover the emergent physics of the $3J$ model beyond the reach of any single method. In particular, it provides an unbiased route to establishing the quantum phase diagram, suggests a gapless intermediate QSL phase, and clarifies the connection of this phase to neighboring ordered phases and to the experimentally observed finite-temperature behavior.

\newsect{Model and Method} The Hamiltonian of kagome antiferromagnetic $3J$ model is
\begin{equation}
H = J_1 \sum_{\langle i,j \rangle_1} \mathbf{S}_i \cdot \mathbf{S}_j
+ J_2 \sum_{\langle i,j \rangle_2} \mathbf{S}_i \cdot \mathbf{S}_j
+ J_{\hexagon} \sum_{\langle i,j \rangle_{\hexagon}} \mathbf{S}_i \cdot \mathbf{S}_j ,
\label{eq:model}
\end{equation}
where $\mathbf{S}_i$ denotes the spin-$1/2$ operator at site $i$ (white dots in the inset of Fig.~\ref{PDDMRG}). All three couplings are antiferromagnetic, $J_1,J_2,J_{\hexagon}>0$, and correspond to nearest-neighbor interactions. To determine its ground-state phase diagram, we employ complementary approaches based on DMRG and NQS. We set $J_{\hexagon}=1$ as the energy unit and focus on the parameter regime $J_1,J_2\in[0,1]$.
For the DMRG calculations, we consider YC cylindrical geometries with a fixed circumference of $L_y=3$ (YC6) and system lengths $L_x=6$, $12$ and $24$, compatible with the 27-site magnetic unit cell of the $3J$ model. 
%\syl{The unit cell is 9-site ?}
These correspond to total spin numbers $N=3\times L_x\times L_y = 54$, $108$ and $216$, respectively. We keep the maximum bond dimension up to $D=15,000$ with $U(1)$ symmetry implemented, achieving well-converged results with truncation errors below $ 1.0\times{10}^{-8}$. 
For the NQS calculations, we adopt a Vision Transformer (ViT) architecture with factored attention~\cite{Viteritti2023, Viteritti2025,Rende2025_QK}, implemented in two variants that differ in their enforcement of translation symmetry. 
%\LuXin{On the $J_1=1$ line, the network enforces the full translation symmetry of the $3J$ Hamiltonian, while on the $J_1=0.5$ line it enforces translation symmetry at the level of a $3\times 3$ supercell of kagome unit cells, which allows for the $(1/3,1/3)$ magnetic order found in the classical phase diagram~\cite{HeringM22}  [cf. Fig.~\ref{fig:arch}]. LuXin: Andreas, please move this detail to the SM part, and explain the rationality for using different strategies for $J_1=1$ and $0.5$.} 
These simulations are performed on an $N=3\times 6\times 6$ system with periodic boundary conditions, and the network parameters are optimized by variational Monte Carlo~\cite{Becca_2017}. %\LuXin{using the NetKet library \cite{netket3:2022,netket2:2019,jax2018github,flax2020github} Please move to the end of Acknowledgments}.  
For finite-temperature properties, we employ the advanced tensor-network methods XTRG and tanTRG as complementary approaches for cross-validation. Both methods are applied to YC6 cylinders with a maximum bond dimension of $D=2000$, which yields well-converged results. Further technical details of the algorithms are provided in the Supplemental Material (SM)~\cite{supp}.

\begin{figure}[htp!]
\includegraphics[width=\columnwidth,angle=0]{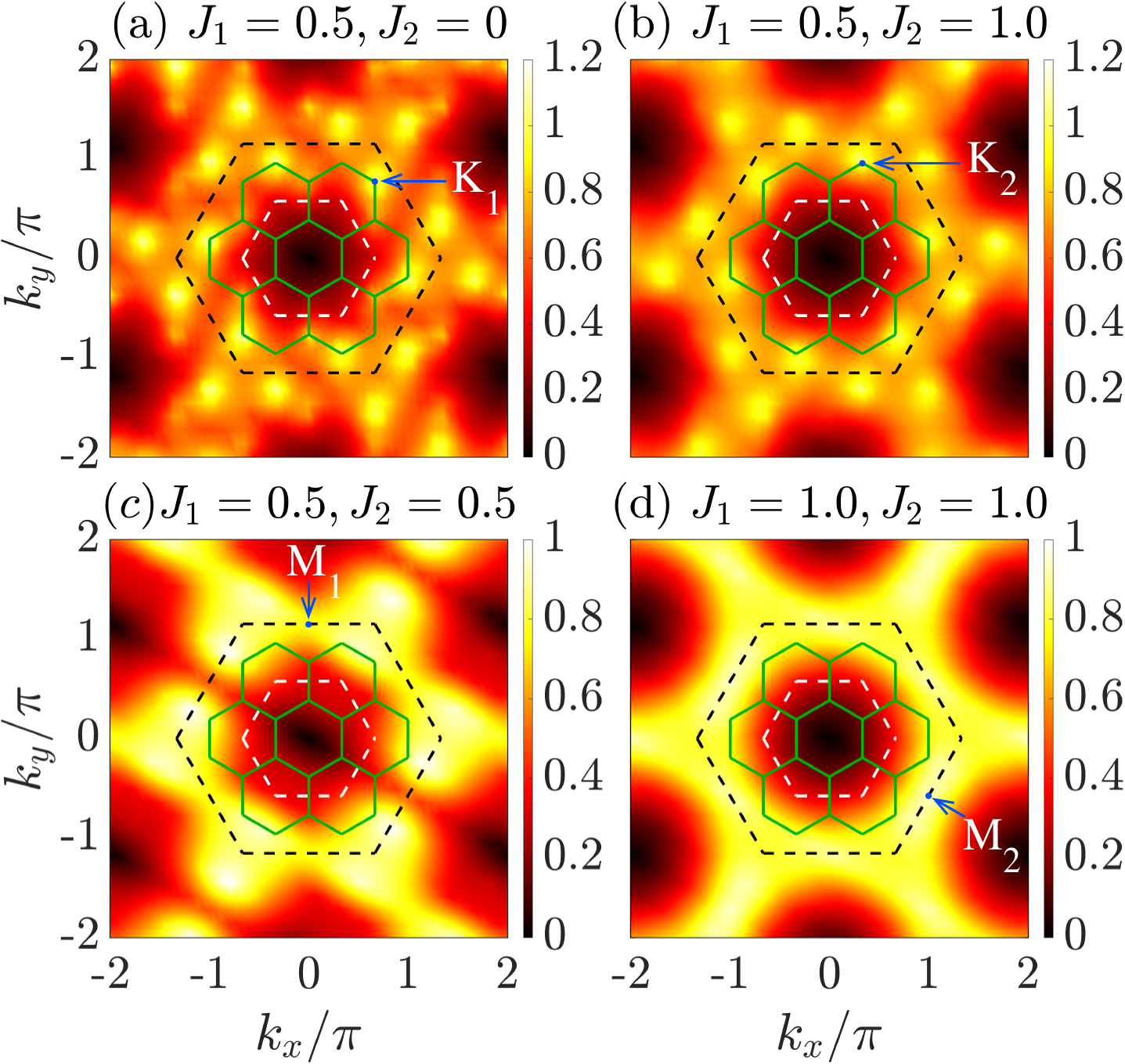}
   \caption{
   \label{SK}
  \textbf{Static spin structure factors.} Representative static spin structure factors $S(\mathbf{k})$ in the (a,b) $(1/3,1/3)$ ordered phases, (c) QSL phase, and (d) KSL phase. The ordering peaks in (a) occur at $\mathbf{Q}=\mathbf{K}_1$, whereas those in (b) occur at $\mathbf{Q}=\mathbf{K}_2$. The positions of ${\mathbf{M}}_1$ and ${\mathbf{M}}_2$ are also denoted directly in (c) and (d). The black dashed hexagon denotes the Brillouin zone (BZ) of the triangular lattice (1-site unit cell), the white dashed hexagon represents the BZ of the kagome lattice (3-site unit cell), and the green solid hexagon corresponds to the BZ of the $3J$ model (9-site unit cell). Here, all $S(\mathbf{k})$ are obtained by Fourier transformation of all-to-all spin-spin correlations. The corresponding NQS results on the $6\times 6$ torus are illustrated in Fig.~\ref{SkNQS} of the SM~\cite{supp}}
\end{figure}

\begin{figure}[htp!]
\includegraphics[width=\columnwidth,angle=0]{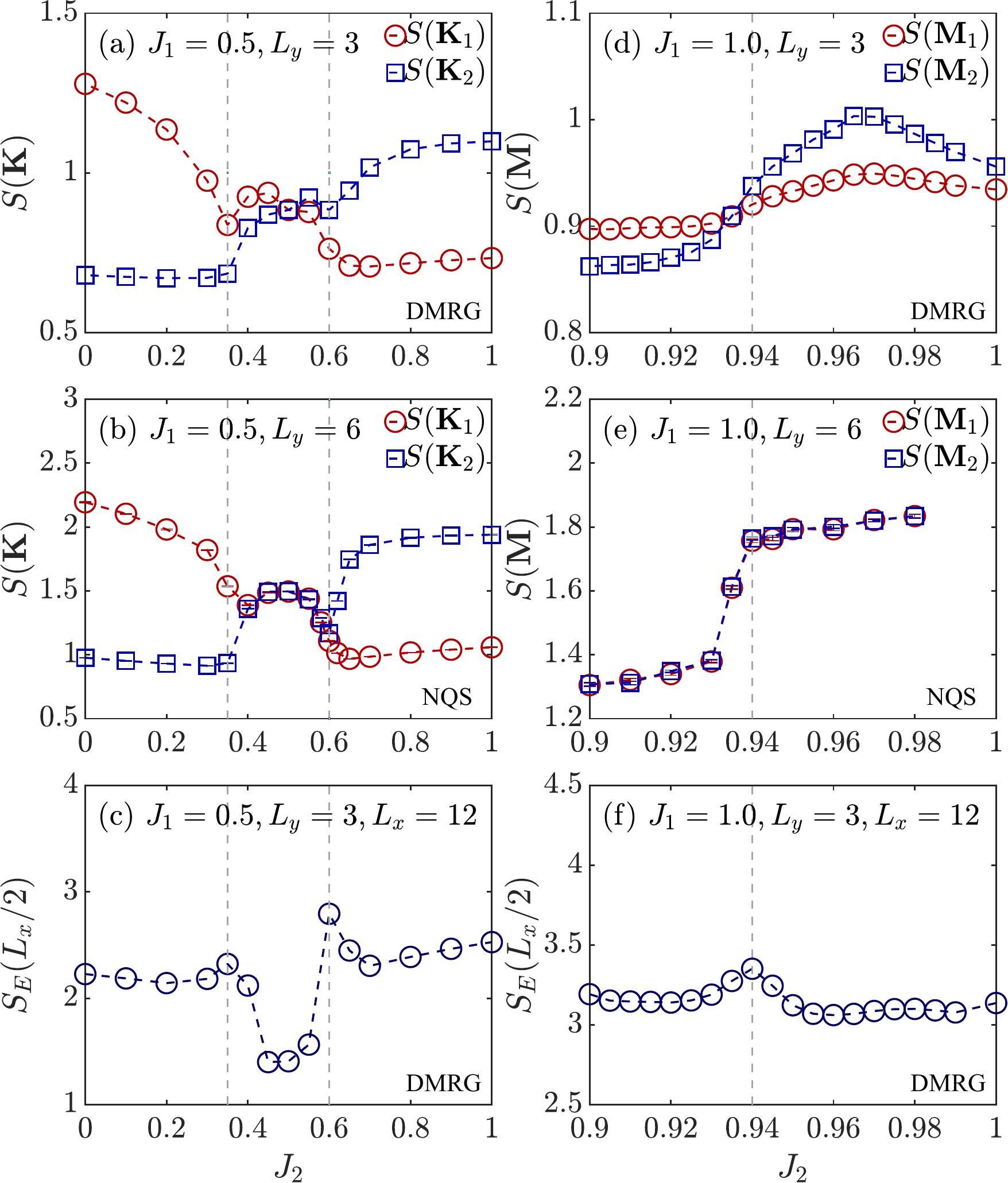}
\caption{\textbf{$J_2$ dependence of $S(\mathbf{k})$ and entanglement entropy.} (a) $J_2$ dependence of $S(\mathbf{k})$ at ${\mathbf{K}}_1$ and ${\mathbf{K}}_2$ for $J_1=0.5$ on YC6 ($L_y=3, L_x=12$) cylinders obtained by DMRG. The kink-like features near $J_2\simeq0.35$ and $0.6$, indicated by the gray dashed lines, signal phase transitions. (b) Same as (a) but on a torus ($L_y=6, L_x=6$) obtained from NQS calculations. (c) $J_2$ dependence of the entanglement entropy $S_{E}(x)$ at the bulk of cylinders $x=L_{x}/2$, yielding critical points consistent with those identified in (a) and (b). (d-f) Same as (a-c), respectively, but for $J_1=1.0$, with $S(\mathbf{k})$ evaluated at ${\mathbf{M}}_1$ and ${\mathbf{M}}_2$, where the phase transition occur near $J_2\simeq0.94$.}
\label{PhaseTransition}
\end{figure}

\newsect{Phase Diagram} Our ground-state results for the $3J$ model obtained from DMRG and NQS calculations are summarized in the phase diagram Fig.~\ref{PDDMRG}, where $J_1$ and $J_2$ are varied with $J_{\hexagon}=1$ fixed. We identify a candidate QSL phase (light-blue region) sandwiched between two $(1/3,1/3)$ ordered phases~\cite{HeringM22} (light-pink regions). Compared with the classical phase boundaries (gray solid lines) determined by classical Monte Carlo calculations~\cite{HeringM22}, the QSL region is reduced in the large-$J_{1,2}$ regimes due to enhanced quantum fluctuations. Moreover, it is concentrated around the diagonal symmetry axis $J_{1}=J_{2}$, where the competing interactions are more balanced in strength, favoring the emergence of the QSL. Notably, we have evidence that 
this QSL phase is separated from the KSL ground state (light-purple regime) in the vicinity of the isotropic $1J$ model by a phase transition. These results imply that distorted kagome $3J$ materials are likely to host two distinct QSL phases.

We characterize these distinct quantum phases by examining the static spin structure factor $S(\mathbf{k})=(1/N)\sum_{i,j}\langle \mathbf{S}_i\cdot \mathbf{S}_j\rangle e^{i\mathbf{k}\cdot(\mathbf{r}_i-\mathbf{r}_j)}$. 
In the $(1/3,1/3)$ ordered phases, $S(\mathbf{k})$ show Bragg peaks at either the $\mathbf{K}_1$ points [cf. Fig.~\ref{SK}(a)] or the $\mathbf{K}_2$ points [cf. Fig.~\ref{SK}(b)]. It is worth pointing out that these two $(1/3,1/3)$ ordered phases are physically equivalent, as both correspond to the same ordering wave vector $\mathbf{Q}=(1/3,1/3)$. The apparent distinction between $\mathbf{K}_1$ and $\mathbf{K}_2$ arises from a transformation that effectively swaps $J_1$ and $J_2$ in the Hamiltonian. In stark contrast, the QSL phase exhibits no sharp Bragg peaks in $S(\mathbf{k})$ [cf. Fig.~\ref{SK}(c)], but rather a broad and featureless distribution, reflecting the absence of long-range magnetic order.  For comparison, we also present $S(\mathbf{k})$ for the KSL phase in Fig.~\ref{SK}(d), which likewise exhibits obvious disordered behavior. However, the topology of $S(\mathbf{k})$ in these two spin liquid phases differs significantly: $S(\mathbf{k})$ is discontinuous at ${\mathbf{M}}_2$ points in the QSL phase, whereas it remains continuous at both ${\mathbf{M}}_1$ and ${\mathbf{M}}_2$ points in the KSL phase. We have verified this distinction across all parameter points within both the QSL and KSL regimes. On the other hand, the real-space spin correlations provide another clear distinction between the QSL and KSL phases [cf. Fig.~\ref{SMSpin} in the SM~\cite{supp}].  In the QSL phase, the spin correlations decay algebraically as a power law, consistent with the behavior expected for a Dirac QSL state, whereas those in the KSL phase decay exponentially. 
Here, we refrain from drawing conclusions about whether the KSL phase is gapped or gapless in the two-dimensional limit, as our results are based solely on finite-size simulations on $L_y=3$ cylinders and an $L_y=6$ torus, while determining the nature of the KSL lies beyond the scope of the present work. Instead, these results provide an additional distinction between the QSL and KSL phases without presupposing the gap structure of the KSL.

To accurately locate the phase boundaries, we track the evolution of $S(\mathbf{k})$ with $J_1$ and $J_2$ at the ordering momenta $\mathbf{K}_1$ and $\mathbf{K}_2$. As a representative cut through the phase diagram, we fix $J_1=0.5$ and vary $J_2$, thereby traversing the $(1/3,1/3)$ ordered and QSL phases [cf. Fig.~\ref{PhaseTransition}(a)]. As $J_2$ increases, the dominant $S(\mathbf{K}_1)$ is gradually suppressed. Around $J_{2}\simeq 0.35$, it exhibits a pronounced kink, accompanied by the gradual development of $S(\mathbf{K}_2)$, signaling a phase transition. In the intermediate regime $0.35 < J_2 < 0.6$, neither $S(\mathbf{K}_1)$ nor $S(\mathbf{K}_2)$ dominates, and both magnetic correlations are substantially suppressed, supporting the presence of a QSL phase. Upon further increasing $J_2$, $S(\mathbf{K}_1)$ is completely suppressed near $J_{2}\simeq 0.6$, while $S(\mathbf{K}_2)$ develops a pronounced kink and becomes the dominant magnetic ordering tendency. Since the computational cost of DMRG grows exponentially with cylinder width $L_y$, we use NQS as a complementary approach to access wider systems on a torus with $L_y=6$ [Fig.~\ref{PhaseTransition}(b)]. Remarkably, the locations of the transition points remain stable against finite-size effects and boundary conditions, further supporting the robustness of the identified phase boundaries. In Fig.~\ref{PhaseTransition}(c), we also show the evolution of the entanglement entropy $S_E(x)=-\mathrm{Tr}\left\lbrack \rho_x \ln\rho_x \right\rbrack$ from DMRG at the middle of the cylinders $x=L_{x}/2$, where $\rho_x$ is the reduced density matrix of the subsystem with rung number $x$. Two kinks are clearly observed, whose locations are also consistent with those identified from Figs.~\ref{PhaseTransition}(a-b). Moreover, the entanglement entropy is larger in the ordered phase, consistent with the expected contribution from Goldstone modes~\cite{VBEE2007Alet}. Additionally, we characterize the QSL–KSL transition by tracking the evolution of $S(\mathbf{k})$ at $\mathbf{M}_1$ and $\mathbf{M}_2$ along a cut with fixed $J_1=1$. Beyond the distinct $S(\mathbf{k})$ profiles shown in Figs.~\ref{SK}(c,d), the QSL–KSL transition is signaled by pronounced kinks in $S(\mathbf{M}_1)$ and $S(\mathbf{M}_2)$ near $J_2\simeq0.94$ [cf. Figs.~\ref{PhaseTransition}(d,e)], accompanied by singular behavior in $S_E(x)$ [cf. Fig.~\ref{PhaseTransition}(f)]. As an additional cross-check of the phase boundaries, we examine the evolution of $S(\mathbf{M})$ in the $(1/3,1/3)$ ordered phases and $S(\mathbf{K})$ in the QSL and KSL phases [cf. Fig.~\ref{PTDMRG} in the SM~\cite{supp}]. The transition points extracted from these independent diagnostics are in excellent agreement.

Finally, to further examine whether the ground state may host VBC states, we also analyze the bond energies across the phase diagram. In both the QSL and KSL phases, we find no discernible signatures of lattice-symmetry breaking, further ruling out VBC as the underlying ground-state order in these regimes [cf. Fig.~\ref{SMVBC} in the SM~\cite{supp}].

\begin{figure}[htp!]
	\centering
	\includegraphics[width=\columnwidth]{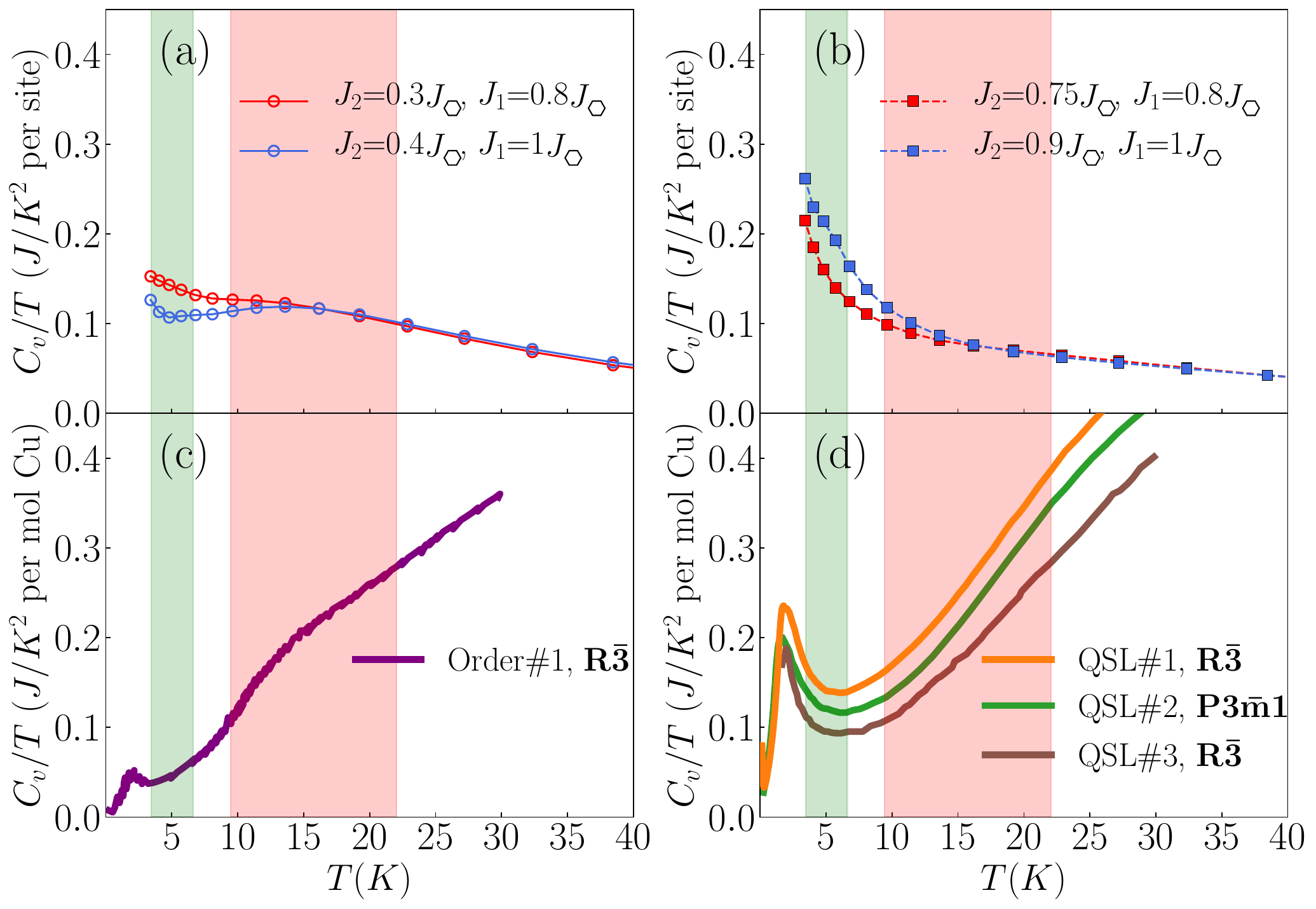}
	\caption{\textbf{Computed and experimental specific heat.} (a) and (b) Calculated $C_v/T$ from XTRG for representative parameter sets in the $(1/3,1/3)$ ordered and QSL phases, respectively. The temperature scale is set by $J_{\hexagon}=63$~K following the experimental estimate~\cite{one2024jeon}. (c) Experimental $C_v/T$ for Y$_3$Cu$_9$(OH)$_{19}$Cl$_8$ from Ref.~\cite{Magneticordering2021Sun} (ordered, marked as Order\#1), which crystallizes in the $R\bar{3}$ space group and realizes the $(1/3,1/3)$-ordered phase. (d) Experimental $C_v/T$ for three candidate QSL materials: disordered LuCu$_3$(OH)$_6$Br$_2$[Br$_x$(OH)$_{1-x}$] in the $R\bar{3}$ space group (orange, QSL\#1)~\cite{Antiferromagnetic2025Zezhong}; YCu$_3$(OH)$_6$Br$_2$[Br$_x$(OH)$_{1-x}$] (green, QSL\#2), which crystallizes in the $P\bar{3}m1$ space group but is also considered a $3J$-model QSL candidate ~\cite{Possibledirac2022Zeng,Spin2025Han}; and YCu$_3$(OH)$_6$O$_{1/3}$Cl$_{8/3}$ (brown, QSL\#3)~\cite{Localstudy2019Barth}, which crystallizes in the $R\bar{3}$ space group.}
	\label{fig:specific_heat}
\end{figure}

\newsect{Specific Heat} To establish a direct connection between the ground states and experimentally accessible thermal signatures, we calculate the specific heat using XTRG and tanTRG, and present the results as $C_v/T$. We consider four representative parameter sets spanning the $(1/3,1/3)$ ordered and QSL phases: $(J_1,J_2)=(0.8,0.3)J_{\hexagon}$ and $(1,0.4)J_{\hexagon}$ in the $(1/3,1/3)$ ordered phase; $(0.8,0.75)J_{\hexagon}$ and $(1,0.85)J_{\hexagon}$ in the QSL phase. The corresponding numerical results are shown in Figs.~\ref{fig:specific_heat}(a-b), with the temperature scale set by $J_{\hexagon}=63$~K following the experimental estimate~\cite{one2024jeon}. We compare these results with experimental data for the ordered material Y$_3$Cu$_9$(OH)$_{19}$Cl$_8$ [cf. Fig.~\ref{fig:specific_heat}(c)] and three candidate QSL materials, including LuCu$_3$(OH)$_6$Br$_2$[Br$_x$(OH)$_{1-x}$], YCu$_3$(OH)$_6$O$_{1/3}$Cl$_{8/3}$), and YCu$_3$(OH)$_6$Br$_2$[Br$_x$(OH)$_{1-x}$] [cf. Fig.~\ref{fig:specific_heat}(d)]. 
We note that the experimental $C_v/T$ data contain a nonmagnetic phonon contribution that is difficult to subtract reliably at low temperatures, especially below $25$~K. Despite this limitation, the experimental data exhibit clear temperature-dependent features that allow a meaningful comparison with our numerical results, revealing two distinct thermal signatures that differentiate the $(1/3,1/3)$-ordered and QSL phases.

First, the $(1/3,1/3)$ ordered phase exhibits a shoulder-like anomaly in $C_v/T$ at $T\sim 10$--$22$~K (red shading), which is absent in the QSL phase. This shoulder consistently appears around $T\sim 15$~K in the ordered compound Y$_3$Cu$_9$(OH)$_{19}$Cl$_8$, but is absent in all three candidate QSL materials. This difference suggests that the shoulder serves as a signature of the zero-temperature magnetic order and thus provides a thermal diagnostic for distinguishing the two phases. Second, at $T\sim 5$~K (green shading), $C_v/T$ in the ordered phase exhibits a pronounced minimum, reaching $\sim 0.1$, while it remains larger ($\sim 0.2$) in the QSL phase. The experimental data show the same trend, with a deep low-temperature dip in the ordered compound and enhanced $C_v/T$ in the QSL candidates. This enhancement implies a larger low-energy density of states in the QSL phase, which naturally accounts for the stronger low-temperature contribution.

\begin{figure}[htp!]
	\centering
	\includegraphics[width=\columnwidth]{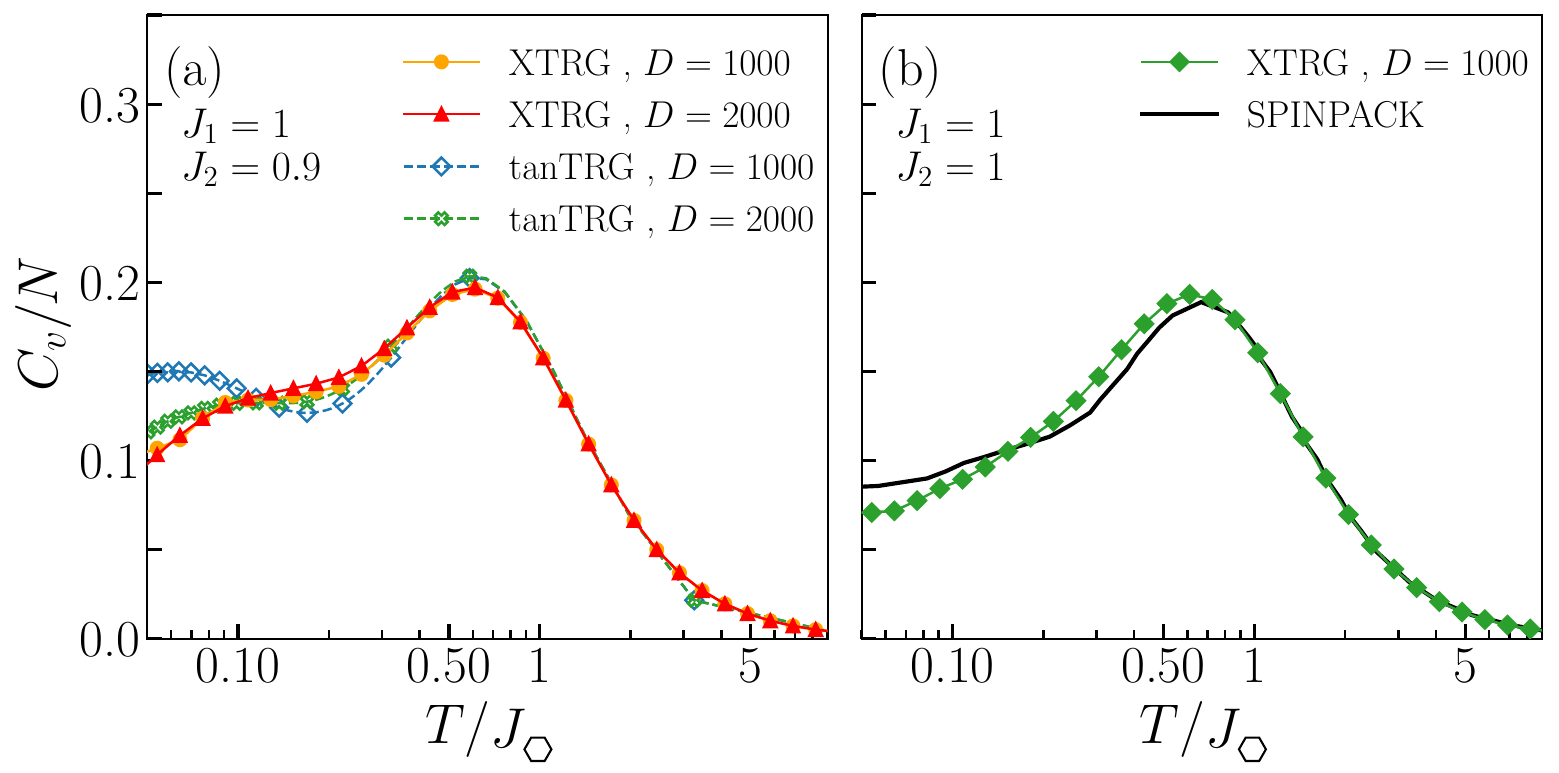}
	\caption{\textbf{Specific heat of the 3$J$ and 1$J$ models.} (a) Calculated specific heat density $C_v/N$, from XTRG and tanTRG caluations with bond dimension $D=1000$ and $2000$. (b) $C_v/N$ at the isotropic point $J_1=J_2=J_{\hexagon}$, from the XTRG calculations with $D=1000$. The black line shows the corresponding $C_v/N$ from SPINPACK calculations reported in Ref.~\cite{Magnetism2018Schnack}.
	}
	\label{fig:specific_heat_org}
\end{figure}

Moreover, in Fig.~\ref{fig:specific_heat_org}(a), we compare XTRG and tanTRG in the QSL phase at $(J_1,J_2)=(1,0.9)J_{\hexagon}$ with bond dimensions up to $D=2000$ to assess the robustness of our numerical results. The XTRG results are well converged for $D=1000$ and $2000$ over the entire temperature range $T/J_{\hexagon} \in [0.05,10]$. In contrast, tanTRG exhibits pronounced $D$ dependence at low temperatures, converging to the XTRG results only as $D$ increases. This confirms the different performance of XTRG and tanTRG and supports the reliability of the results in Figs.~\ref{fig:specific_heat}(a-b).
On the other hand, we also compare our XTRG results at the isotropic $1J$ kagome limit with SPINPACK data from Ref.~\cite{Magnetism2018Schnack} [cf. Fig.~\ref{fig:specific_heat_org}(b)]. The 
good
agreement further validates our approach.
Finally, to highlight the distinct thermal behavior, we compare $C_v/N$ at $(J_1,J_2)=(1.0,0.9)J_{\hexagon}$ with that of the $1J$ model. The QSL-phase data exhibit a prominent shoulder around $T/J_{\hexagon}\sim0.1$, whereas the $1J$ model decreases monotonically and smoothly over the same range. This striking difference corroborates our ground-state phase diagram analysis, providing strong evidence that the QSL phase of the $3J$ model is not adiabatically connected to the $1J$ limit.

\newsect{Summary and Discussion} In this work, we have established the ground-state and finite-temperature properties of the kagome antiferromagnetic $3J$ model through an integrated investigation, combining two complementary ground-state methods (DMRG and NQS), two finite-temperature tensor-network approaches (XTRG and tanTRG), and experimental specific-heat data from four $3J$ materials. These independent approaches consistently identify an intermediate QSL phase between two $(1/3,1/3)$ ordered phases and establish that it is distinct from the KSL phase at the isotropic $1J$ limit. The consistency between DMRG and NQS results, despite their different geometries and effective system sizes, mitigates uncertainties associated with boundary conditions and finite-size effects, rendering the identification of the QSL phase particularly robust and compelling. The rich quantum phase diagram also highlights the intricate physics arising from the interplay of inequivalent exchange couplings. The calculated specific heat further captures the characteristic shoulder in the ordered regime and its disappearance in the QSL regime, in close agreement with experiments on both ordered and QSL samples.

Beyond establishing the phase diagram and finite-temperature signatures, our integrated analysis also raises several important questions concerning the low-energy physics of the QSL phase. Resolving the corresponding low-energy Dirac-cone structure in the excitation spectrum using both analytic~\cite{proximate2024Scheie} and numerical~\cite{Hu_DSL_PRL2019,Jiang_DDMRG_2026,Xiang_PRL2026,Pollmann_PRB2023} approaches would provide a more direct characterization of its underlying field theory. On the experimental side, we note that our thermal measurements have not yet accessed a sufficiently low-temperature regime to resolve the $T^2$ specific heat ($C_v \propto T^2$) expected for a Dirac QSL~\cite{song2026ss,guoTlinear2026,RanY07}. Accessing this regime numerically will require further advances in finite-temperature numerical methods to enable reliable simulations at lower temperatures. Besides, the magnetic field response~\cite{Bader_PRB2026,song2026ss,Starykh2025} of the $3J$ model also warrants investigation, including the intriguing $1/9$ magnetization plateau and quantum oscillations observed therein~\cite{Guoxin_PRX2025,Kumar_PRL2026,Byungmin_2025,Unconventionalmagnetic2025Zheng,one2024jeon}, as well as possible field-induced phase transitions, which may provide further insight into the QSL-KSL transition and offer additional probes of the underlying low-energy theory.

%In parallel,  NQS have demonstrated  their ability to accurately capture ground states, low-lying excitations, and phase boundaries in frustrated magnets, including the square- and triangular-lattice $J_1$-$J_2$ models~\cite{Roth_PRB2023, Nomura2021} and the Shastry-Sutherland model~\cite{Viteritti2025,Neural2023Mezera,Design2025Nutakki}. Recent studies using variational wave functions based on the transformer architecture have reached periodic  clusters of several hundred to more than a thousand spins on the square and triangular lattices~\cite{Rende2026scaling,Viteritti2026thermolimit}, and have enabled controlled finite-size extrapolations of energies and order parameters ~\cite{Viteritti2026thermolimit}.  The $3J$ model thus provides a natural yet more demanding setting for such scaling efforts. Its enlarged unit cell and inequivalent exchange couplings lead to a richer low-energy spectrum and substantially more intricate momentum-space structure, while its close connection to experimentally accessible materials makes quantitative theoretical predictions especially valuable. Extending NQS to larger kagome clusters~\cite{raikosVariational2026}, excited-state spectra, and finite-temperature observables such as the specific heat could therefore offer an independent route to the low-energy and thermodynamic properties of the $3J$ model, complementary to conventional tensor-network approaches.

In parallel, NQS approaches have demonstrated their ability to accurately capture ground states, low-lying excitations, and phase boundaries in frustrated magnets, including the square- and triangular-lattice $J_1$-$J_2$ models~\cite{Roth_PRB2023,Nomura2021} and the Shastry-Sutherland model~\cite{Viteritti2025,Neural2023Mezera,Design2025Nutakki}. Recent studies using variational wave functions based on the transformer architecture have reached periodic clusters of several hundred to more than a thousand spins on the square and triangular lattices~\cite{Rende2026scaling,Viteritti2026thermolimit}, enabling controlled finite-size extrapolations of energies and order parameters~\cite{Viteritti2026thermolimit}. The $3J$ model provides a natural yet demanding setting for such scaling efforts, owing to its enlarged unit cell, inequivalent exchange couplings and frustrated geometry, while its close connection to experimentally accessible materials makes quantitative theoretical predictions especially valuable. Extending NQS simulations to magnetization processes~\cite{raikosVariational2026}, excited-state spectra, and finite-temperature observables such as the specific heat~\cite{Nomura2021_thermal} could therefore offer an independent route to the in-field, low-energy and thermodynamic properties of the $3J$ model, complementary to conventional tensor-network approaches.

\newsect{Acknowledgments} We thank Tim-Lok Chau, Min Long and Xu-Yan Jia for helpful comments on the manuscript and for the suggested references. We thank Fabien Alet for inspiring discussions on the neural quantum states. XL, MHS, ZYM and CKZ acknowledge the support from the Research Grants Council (RGC) of Hong Kong (Project Nos. C7037-22GF, 17302223, 17301924, 17301725, 17302026) and the State Key Laboratory of Optical Quantum Materials at HKU, and additionally with SC the ANR/RGC Joint Research Scheme sponsored by RGC of Hong Kong (Project No. A\_HKU703/22) and the French National Research Agency (grant ANR-22-CE30-0042-01).  We thank the HPC2021 system under the Information Technology Services at the University of Hong Kong~\cite{hpc2021}, as well as the Beijing Paratera Tech Corp., Ltd~\cite{paratera}, CALMIP (grant 2025-P0677 on Olympe cluster, project M25154 on Turpan cluster) and GENCI (project A0190500225), for providing HPC resources that have contributed to the research results reported within this paper. LS acknowledge the support from the National Key Research and Development Program of China (Grants No. 2022YFA1403400, No. 2021YFA1400400). 
%\LuXin{The tensor network calculations in this work are performed using TensorKit library~\cite{TensorKit}.}
The tensor network calculations in this work are performed using TensorKit library~\cite{TensorKit} while the NQS simulations are performed using NetKet~\cite{netket2:2019,netket3:2022}, which is built on top of the JAX~\cite{jax2018github} and Flax~\cite{flax2020github} libraries.

\newsect{Data availability} The data are available from the authors upon reasonable request.

\bibliography{ref.bib}

@article{proximate2024Scheie,
	title = {{Proximate spin liquid and fractionalization in the triangular antiferromagnet KYbSe$_2$}},
	author = {Scheie, A. O. and Ghioldi, E. A. and Xing, J. and Paddison, J. A. M. and Sherman, N. E. and Dupont, M. and Sanjeewa, L. D. and Lee, Sangyun and Woods, A. J. and Abernathy, D. and Pajerowski, D. M. and Williams, T. J. and Zhang, Shang-Shun and Manuel, L. O. and Trumper, A. E. and Pemmaraju, C. D. and Sefat, A. S. and Parker, D. S. and Devereaux, T. P. and Movshovich, R. and Moore, J. E. and Batista, C. D. and Tennant, D. A.},
	journal = {Nat. Phys.},
	volume = {20}, 
	pages = {74--81},
	year = {2024},
	doi = {10.1038/s41567-023-02259-1},
    url = {https://doi.org/10.1038/s41567-023-02259-1}
}

@article{VBEE2007Alet,
  title = {Valence Bond Entanglement Entropy},
  author = {Alet, Fabien and Capponi, Sylvain and Laflorencie, Nicolas and Mambrini, Matthieu},
  journal = {Phys. Rev. Lett.},
  volume = {99},
  issue = {11},
  pages = {117204},
  numpages = {4},
  year = {2007},
  month = {Sep},
  publisher = {American Physical Society},
  doi = {10.1103/PhysRevLett.99.117204},
  url = {https://link.aps.org/doi/10.1103/PhysRevLett.99.117204}
}

@article{Neural2023Mezera,
	title = {{Neural network quantum states analysis of the Shastry-Sutherland model}},
	pages = {088},
	author = {Mezera, Matej and Menšíková, Jana and Baláž, Pavel and Žonda, Martin},
	journal = {SciPost Phys. Core},
	volume = {6},
	year = {2023},
	publisher = {SciPost},
	doi = {10.21468/SciPostPhysCore.6.4.088},
	url = {https://scipost.org/10.21468/SciPostPhysCore.6.4.088}
}

@article{Design2025Nutakki,
  title = {{Design principles of deep translationally symmetric neural quantum states for frustrated magnets}},
  author = {Nutakki, Rajah P. and Shokry, Ahmedeo and Vicentini, Filippo},
  journal = {Phys. Rev. Res.},
  volume = {7},
  issue = {4},
  pages = {043099},
  numpages = {13},
  year = {2025},
  month = {Oct},
  publisher = {American Physical Society},
  doi = {10.1103/ybgv-35jm},
  url = {https://link.aps.org/doi/10.1103/ybgv-35jm}
}

@article{Li2025Recent,
doi = {10.1088/0256-307X/42/7/070716},
url = {https://doi.org/10.1088/0256-307X/42/7/070716},
year = {2025},
month = {jul},
publisher = {Chinese Physical Society and IOP Publishing Ltd},
volume = {42},
number = {7},
pages = {070716},
author = {Li, Shiliang},
title = {{Recent Advances in Quantum Spin Liquids in Two-Dimensional Kagome System $\mathrm{YCu}_3 \mathrm{(OH)}_{6+x}\mathrm{X}_{3-x} (\mathrm{X} = \mathrm{Cl, Br})$}},
journal = {Chin. Phys. Lett.}
}

@article{abbruciati2026ab,
  title={{Ab Initio Investigation of Pressure Effects in the Spin-Liquid Candidate Y-Kapellasite}},
  author={Abbruciati, Federico and Razpopov, Aleksandar and Rodrigues, Joao Elias FS and Garbarino, Gaston and Tacon, Matthieu Le and Valenti, Roser and Puphal, Pascal and Wehinger, Bj{\"o}rn},
  journal={arXiv preprint arXiv:2607.07341},
  year={2026},
  url={https://arxiv.org/pdf/2607.07341}
}

@article{Liu2023Tangent,
  title = {{Tangent Space Approach for Thermal Tensor Network Simulations of the 2D Hubbard Model}},
  author = {Li, Qiaoyi and Gao, Yuan and He, Yuan-Yao and Qi, Yang and Chen, Bin-Bin and Li, Wei},
  journal = {Phys. Rev. Lett.},
  volume = {130},
  issue = {22},
  pages = {226502},
  numpages = {8},
  year = {2023},
  month = {Jun},
  publisher = {American Physical Society},
  doi = {10.1103/PhysRevLett.130.226502},
  url = {https://link.aps.org/doi/10.1103/PhysRevLett.130.226502}
}

@article{Localstudy2019Barth,
  title = {{Local study of the insulating quantum kagome antiferromagnets ${\mathrm{YCu}}_{3}{(\mathrm{OH})}_{6}{\mathrm{O}}_{x}{\mathrm{Cl}}_{3\ensuremath{-}x}(x=0,1/3)$}},
  author = {Barth\'elemy, Quentin and Puphal, Pascal and Zoch, Katharina M. and Krellner, Cornelius and Luetkens, Hubertus and Baines, Christopher and Sheptyakov, Denis and Kermarrec, Edwin and Mendels, Philippe and Bert, Fabrice},
  journal = {Phys. Rev. Mater.},
  volume = {3},
  issue = {7},
  pages = {074401},
  numpages = {9},
  year = {2019},
  month = {Jul},
  publisher = {American Physical Society},
  doi = {10.1103/PhysRevMaterials.3.074401},
  url = {https://link.aps.org/doi/10.1103/PhysRevMaterials.3.074401}
}

@article{Magnetism2018Schnack,
  title = {{Magnetism of the $N=42$ kagome lattice antiferromagnet}},
  author = {Schnack, J\"urgen and Schulenburg, J\"org and Richter, Johannes},
  journal = {Phys. Rev. B},
  volume = {98},
  issue = {9},
  pages = {094423},
  numpages = {10},
  year = {2018},
  month = {Sep},
  publisher = {American Physical Society},
  doi = {10.1103/PhysRevB.98.094423},
  url = {https://link.aps.org/doi/10.1103/PhysRevB.98.094423}
}

@article{one2024jeon,
  title={{One-ninth magnetization plateau stabilized by spin entanglement in a kagome antiferromagnet}},
  author={Jeon, Sungmin and Wulferding, Dirk and Choi, Youngsu and Lee, Seungyeol and Nam, Kiwan and Kim, Kee Hoon and Lee, Minseong and Jang, Tae-Hwan and Park, Jae-Hoon and Lee, Suheon and others},
  journal={Nat. Phys.},
  volume={20},
  number={3},
  pages={435--441},
  year={2024},
  publisher={Nature Publishing Group UK London},
  url={https://doi.org/10.1038/s41567-023-02318-7}
}

@article{Trotter1959,
 ISSN = {00029939, 10886826},
 URL = {http://www.jstor.org/stable/2033649},
 author = {H. F. Trotter},
 journal = {Proceedings of the American Mathematical Society},
 number = {4},
 pages = {545--551},
 publisher = {American Mathematical Society},
 title = {On the Product of Semi-Groups of Operators},
 urldate = {2026-02-03},
 volume = {10},
 year = {1959}
}

@article{Suzuki1990,
title = {{Fractal decomposition of exponential operators with applications to many-body theories and Monte Carlo simulations}},
journal = {Physics Letters A},
volume = {146},
number = {6},
pages = {319-323},
year = {1990},
issn = {0375-9601},
doi = {https://doi.org/10.1016/0375-9601(90)90962-N},
url = {https://www.sciencedirect.com/science/article/pii/037596019090962N},
author = {Masuo Suzuki}
}

@article{Chen2017Series,
  title = {{Series-expansion thermal tensor network approach for quantum lattice models}},
  author = {Chen, Bin-Bin and Liu, Yun-Jing and Chen, Ziyu and Li, Wei},
  journal = {Phys. Rev. B},
  volume = {95},
  issue = {16},
  pages = {161104},
  numpages = {5},
  year = {2017},
  month = {Apr},
  publisher = {American Physical Society},
  doi = {10.1103/PhysRevB.95.161104},
  url = {https://link.aps.org/doi/10.1103/PhysRevB.95.161104}
}

@Article{netket3:2022,
    title={{NetKet} 3: {Machine} Learning Toolbox for Many-Body Quantum Systems},
    author={Filippo Vicentini and Damian Hofmann and Attila Szabó and Dian Wu and Christopher Roth and Clemens Giuliani and Gabriel Pescia and Jannes Nys and Vladimir Vargas-Calderón and Nikita Astrakhantsev and Giuseppe Carleo},
    journal={SciPost Phys. Codebases},
    pages={7},
    year={2022},
    publisher={SciPost},
    doi={10.21468/SciPostPhysCodeb.7},
    url={https://scipost.org/10.21468/SciPostPhysCodeb.7}
}

@article{netket2:2019,
    title={{NetKet: A machine learning toolkit for many-body quantum systems}},
    author={Carleo, Giuseppe and Choo, Kenny and Hofmann, Damian and Smith, James ET and Westerhout, Tom and Alet, Fabien and Davis, Emily J and Efthymiou, Stavros and Glasser, Ivan and Lin, Sheng-Hsuan and Mauri, Marta and Mazzola, Guglielmo and Pereira, Christian B and Vicentini, Filippo},
    journal={SoftwareX},
    volume={10},
    pages={100311},
    year={2019},
    publisher={Elsevier},
    doi={10.1016/j.softx.2019.100311},
    url={https://www.sciencedirect.com/science/article/pii/S2352711019300974}
}

@misc{jax2018github,
  title = {{{{JAX}}: Composable Transformations of {{Python}}+{{NumPy}} Programs}},
  author = {Bradbury, James and Frostig, Roy and Hawkins, Peter and Johnson, Matthew James and Leary, Chris and Maclaurin, Dougal and Necula, George and Paszke, Adam and VanderPlas, Jake and {Wanderman-Milne}, Skye and Zhang, Qiao},
  year = {2018}
}

@misc{flax2020github,
  title = {{Flax: {{A}} Neural Network Library and Ecosystem for {{JAX}}}},
  author = {Heek, Jonathan and Levskaya, Anselm and Oliver, Avital and Ritter, Marvin and Rondepierre, Bertrand and Steiner, Andreas and {van Zee}, Marc},
  year = {2024}
}

@article{Magneticordering2021Sun,
  title = {{Magnetic ordering of the distorted kagome antiferromagnet ${\mathrm{Y}}_{3}{\mathrm{Cu}}_{9}{(\mathrm{OH})}_{18}[{\mathrm{Cl}}_{8}$(OH)] prepared via optimal synthesis}},
  author = {Sun, W. and Arh, T. and Gomil\ifmmode \check{s}\else \v{s}\fi{}ek, M. and Ko\ifmmode \check{z}\else \v{z}\fi{}elj, P. and Vrtnik, S. and Herak, M. and Mi, J.-X. and Zorko, A.},
  journal = {Phys. Rev. Mater.},
  volume = {5},
  issue = {6},
  pages = {064401},
  numpages = {11},
  year = {2021},
  month = {Jun},
  publisher = {American Physical Society},
  doi = {10.1103/PhysRevMaterials.5.064401},
  url = {https://link.aps.org/doi/10.1103/PhysRevMaterials.5.064401}
}

@article{Chen2018Exponential,
title = {Exponential Thermal Tensor Network Approach for Quantum Lattice Models},
author = {Chen, Bin-Bin and Chen, Lei and Chen, Ziyu and Li, Wei and Weichselbaum, Andreas},
journal = {Phys. Rev. X},
volume = {8},
issue = {3},
pages = {031082},
numpages = {29},
year = {2018},
month = {Sep},
publisher = {American Physical Society},
doi = {10.1103/PhysRevX.8.031082},
url = {https://link.aps.org/doi/10.1103/PhysRevX.8.031082}
}

@article{Spectralevidence2024Zeng,
author = {Zeng, Zhenyuan and Zhou, Chengkang and Zhou, Honglin and Han, Lankun and Chi, Runze and Li, Kuo and Kofu, Maiko and Nakajima, Kenji and Wei, Yuan and Zhang, Wenliang and Mazzone, Daniel G. and Meng, Zi Yang and Li, Shiliang},
year = {2024},
title ={{Spectral evidence for Dirac spinons in a kagome lattice antiferromagnet}},
journal = {Nat. Phys.},
pages = {1097 - 1102},
volume = {20},
URL = {https://doi.org/10.1038/s41567-024-02495-z},
doi = {10.1038/s41567-024-02495-z}
}

@article{Unconventionalmagnetic2025Zheng,
author = {Guoxin Zheng  and Yuan Zhu  and Kuan-Wen Chen  and Byungmin Kang  and Dechen Zhang  and Kaila Jenkins  and Aaron Chan  and Zhenyuan Zeng  and Aini Xu  and Oscar A. Valenzuela  and Joanna Blawat  and John Singleton  and Shiliang Li  and Patrick A. Lee  and Lu Li },
title = {{Unconventional magnetic oscillations in a kagome Mott insulator}},
journal = {Proc. Natl. Acad. Sci. U.S.A},
volume = {122},
number = {5},
pages = {e2421390122},
year = {2025},
doi = {10.1073/pnas.2421390122},
URL = {https://www.pnas.org/doi/abs/10.1073/pnas.2421390122},
}

@article{Antiferromagnetic2025Zezhong,
title = {Antiferromagnetic Order and Possible Quantum Spin Liquid in Kagome Antiferromagnet {LuCu$_3$(OH)$_6$Br$_2$[Br$_x$(OH)$_{1–x}$]}},
author = {Zezhong Li and Lankun Han and Yili Sun and Shanshan Zhang and Zhenyuan Zeng and Shiliang Li},
journal = {Chin. Phys. Lett.},
volume = {42},
pages = {027504},
year = {2025},
doi = {10.1088/0256-307X/42/2/027504},	
url = {http://cpl.iphy.ac.cn/en/article/doi/10.1088/0256-307X/42/2/027504},
}

@article{Spin2025Han,
  title = {{Spin excitations arising from anisotropic Dirac spinons in ${\mathrm{YCu}}_{3}{(\mathrm{OD})}_{6}{\mathrm{Br}}_{2}[{\mathrm{Br}}_{0.33}{(\mathrm{OD})}_{0.67}]$}},
  author = {Han, Lankun and Zeng, Zhenyuan and Long, Min and Song, Menghan and Zhou, Chengkang and Liu, Bo and Kofu, Maiko and Nakajima, Kenji and Steffens, Paul and Hiess, Arno and Meng, Zi Yang and Su, Yixi and Li, Shiliang},
  journal = {Phys. Rev. B},
  volume = {112},
  issue = {4},
  pages = {045114},
  numpages = {11},
  year = {2025},
  month = {Jul},
  publisher = {American Physical Society},
  doi = {10.1103/qypv-n4yg},
  url = {https://link.aps.org/doi/10.1103/qypv-n4yg}
}

@article{XuA24,
	title = {{Magnetic ground states in the kagome system ${\mathrm{YCu}}_{3}{(\mathrm{OH})}_{6}[{({\mathrm{Cl}}_{x}{\mathrm{Br}}_{1\ensuremath{-}x})}_{3\ensuremath{-}y}{(\mathrm{OH})}_{y}]$}},
	author = {Xu, Aini and Shen, Qinxin and Liu, Bo and Zeng, Zhenyuan and Han, Lankun and Yan, Liqin and Luo, Jun and Yang, Jie and Zhou, Rui and Li, Shiliang},
	journal = {Phys. Rev. B},
	volume = {110},
	issue = {8},
	pages = {085146},
	numpages = {10},
	year = {2024},
	month = {Aug},
	publisher = {American Physical Society},
	doi = {10.1103/PhysRevB.110.085146},
	url = {https://link.aps.org/doi/10.1103/PhysRevB.110.085146}
}

@article{Possibledirac2022Zeng,
  title = {{Possible Dirac quantum spin liquid in the kagome quantum antiferromagnet ${\mathrm{YCu}}_{3}{(\mathrm{OH})}_{6}{\mathrm{Br}}_{2}[{\mathrm{Br}}_{x}{(\mathrm{OH})}_{1\ensuremath{-}x}]$}},
  author = {Zeng, Zhenyuan and Ma, Xiaoyan and Wu, Si and Li, Hai-Feng and Tao, Zhen and Lu, Xingye and Chen, Xiao-hui and Mi, Jin-Xiao and Song, Shi-Jie and Cao, Guang-Han and Che, Guangwei and Li, Kuo and Li, Gang and Luo, Huiqian and Meng, Zi Yang and Li, Shiliang},
  journal = {Phys. Rev. B},
  volume = {105},
  issue = {12},
  pages = {L121109},
  numpages = {6},
  year = {2022},
  month = {Mar},
  publisher = {American Physical Society},
  doi = {10.1103/PhysRevB.105.L121109},
  url = {https://link.aps.org/doi/10.1103/PhysRevB.105.L121109}
}

@article{FerrariF24,
	title = {{Spin-phonon interactions on the kagome lattice: Dirac spin liquid versus valence-bond solids}},
	author = {Ferrari, Francesco and Becca, Federico and Valent\'{\i}, Roser},
	journal = {Phys. Rev. B},
	volume = {109},
	issue = {16},
	pages = {165133},
	numpages = {8},
	year = {2024},
	month = {Apr},
	publisher = {American Physical Society},
	doi = {10.1103/PhysRevB.109.165133},
	url = {https://link.aps.org/doi/10.1103/PhysRevB.109.165133}
}

@article{ZhuW19,
	title = {{Identifying spinon excitations from dynamic structurefactor of spin-1/2 {Heisenberg} antiferromagnet onthe Kagome lattice}},
	author = {W. Zhu and Shou-shu Gong and D. N. Sheng},
	journal = {Proc. Natl. Acad. Sci. U.S.A},
	volume = {116},
	pages = {5437},
	year = {2019},
	doi = {10.1073/pnas.1807840116}
}

@article{IqbalY21,
	title = {{Gutzwiller projected states for the ${J}_{1}\ensuremath{-}{J}_{2}$ Heisenberg model on the Kagome lattice: Achievements and pitfalls}},
	author = {Iqbal, Yasir and Ferrari, Francesco and Chauhan, Aishwarya and Parola, Alberto and Poilblanc, Didier and Becca, Federico},
	journal = {Phys. Rev. B},
	volume = {104},
	issue = {14},
	pages = {144406},
	numpages = {9},
	year = {2021},
	month = {Oct},
	publisher = {American Physical Society},
	doi = {10.1103/PhysRevB.104.144406},
	url = {https://link.aps.org/doi/10.1103/PhysRevB.104.144406}
}

@article{KieseD23,
	title = {{Pinch-points to half-moons and up in the stars: The kagome skymap}},
	author = {Kiese, Dominik and Ferrari, Francesco and Astrakhantsev, Nikita and Niggemann, Nils and Ghosh, Pratyay and M\"uller, Tobias and Thomale, Ronny and Neupert, Titus and Reuther, Johannes and Gingras, Michel J. P. and Trebst, Simon and Iqbal, Yasir},
	journal = {Phys. Rev. Res.},
	volume = {5},
	issue = {1},
	pages = {L012025},
	numpages = {8},
	year = {2023},
	month = {Feb},
	publisher = {American Physical Society},
	doi = {10.1103/PhysRevResearch.5.L012025},
	url = {https://link.aps.org/doi/10.1103/PhysRevResearch.5.L012025}
}

@article{ZhuW18,
	title = {{Entanglement signatures of emergent Dirac fermions: Kagome spin liquid and quantum criticality}},
	author = {Wei Zhu and Xiao Chen and Yin-Chen He and William Witczak-Krempa},
	journal = {Sci. Adv.},
	volume = {4},
	pages = {eaat5535},
	year = {2018},
	doi = {10.1126/sciadv.aat5535}
}

@article{HeringM22,
	title = {{Phase diagram of a distorted kagome antiferromagnet and application to {Y-kapellasite}}},
	author = "Max Hering and Francesco Ferrari and Aleksandar Razpopov and Igor I. Mazin and Roser Valentí and Harald O. Jeschke and Johannes Reuther",
	journal = "npj Comput. Mater.",
	volume = "8", 
	pages = "10",
	year = "2022",
	doi = "10.1038/s41524-021-00689-0"
}

@article{ChatterjeeD23,
	title = {{From spin liquid to magnetic ordering in the anisotropic kagome Y-kapellasite ${\mathrm{Y}}_{3}{\mathrm{Cu}}_{9}{(\mathrm{OH})}_{19}{\mathrm{Cl}}_{8}$: A single-crystal study}},
	author = {Chatterjee, Dipranjan and Puphal, Pascal and Barth\'elemy, Quentin and Willwater, Jannis and S\"ullow, Stefan and Baines, Christopher and Petit, Sylvain and Ressouche, Eric and Ollivier, Jacques and Zoch, Katharina M. and Krellner, Cornelius and Parzer, Michael and Riss, Alexander and Garmroudi, Fabian and Pustogow, Andrej and Mendels, Philippe and Kermarrec, Edwin and Bert, Fabrice},
	journal = {Phys. Rev. B},
	volume = {107},
	issue = {12},
	pages = {125156},
	numpages = {13},
	year = {2023},
	month = {Mar},
	publisher = {American Physical Society},
	doi = {10.1103/PhysRevB.107.125156},
	url = {https://link.aps.org/doi/10.1103/PhysRevB.107.125156}
}

@article{Dynamicfingerprint2021Fu,
	title = {{Dynamic fingerprint of fractionalized excitations in single-crystalline {Cu$_3$Zn(OH)$_6$FBr}}},
	author = "Ying Fu and Miao-Ling Lin and Le Wang and Qiye Liu and Lianglong Huang and Wenrui Jiang and Zhanyang Hao and Cai Liu and Hu Zhang and Xingqiang Shi and Jun Zhang and Junfeng Dai and Dapeng Yu and Fei Ye and Patrick A. Lee and Ping-Heng Tan and Jia-Wei Mei",
	journal = "Nat. Commun.",
	volume = "12", 
	pages = "3048",
	year = "2021",
	doi = "10.1038/s41467-021-23381-9"
}

@article{WeiY21,
	title = {Nonlocal Effects of Low-Energy Excitations in Quantum-Spin-Liquid Candidate {Cu$_3$Zn(OH)$_6$FBr}},
	author = "Yuan Wei and Xiaoyan Ma and Zili Feng and Yongchao Zhang and Lu Zhang and Huaixin Yang and Yang Qi and Zi Yang Meng and Yan-Cheng Wang and Youguo Shi and Shiliang Li",
	journal = "Chin. Phys. Lett.",
	volume = "38", 
	pages = "097501",
	year = "2021",
	doi = "10.1088/0256-307X/38/9/097501"
}

@misc{supp,
  title = {See Supplemental Materials for more supporting data.}, 
}

@article{VriesMA08,
	title = {Magnetic Ground State of an Experimental {$S=1/2$} Kagome Antiferromagnet},
	author = {de Vries, M. A. and Kamenev, K. V. and Kockelmann, W. A. and Sanchez-Benitez, J. and Harrison, A.},
	journal = {Phys. Rev. Lett.},
	volume = {100},
	issue = {15},
	pages = {157205},
	numpages = {4},
	year = {2008},
	month = {Apr},
	publisher = {American Physical Society},
	doi = {10.1103/PhysRevLett.100.157205},
	url = {https://link.aps.org/doi/10.1103/PhysRevLett.100.157205}
}

@article{FreedmanDE10,
	title = {{Site Specific X-ray Anomalous Dispersion of the Geometrically Frustrated Kagom\'{e} Magnet, Herbertsmithite, {ZnCu$_3$(OH)$_6$Cl$_2$}}},
	author = "Danna E. Freedman and Tianheng H. Han and  Andrea Prodi and Peter M{\"{u}}ller and Qing-Zhen Huang and Yu-Sheng Chen and Samuel M. Webb and Young S. Lee and Tyrel M. McQueen and Daniel G. Nocera",
	journal= "J. Am. Chem. Soc.",
	volume = "132",
	pages = "16185",
	year = "2010",
	doi = "10.1021/ja1070398"
}

@article{KhuntiaP20,
	title = {{Gapless ground state in the archetypal quantum kagome antiferromagnet {ZnCu$_3$(OH)$_6$Cl$_2$}}},
	author = "P. Khuntia and M. Velazquez and Q. Barth\'{e}lemy and F. Bert and E. Kermarrec and A. Legros and B. Bernu and L. Messio, A. Zorko and P. Mendels",
	journal = "Nat. Phys.",
	volume = "16",
	pages = "469",
	year = "2020",
	doi = "10.1038/s41567-020-0792-1"
}

@ARTICLE{HanTH16,
	title = {{Correlated impurities and intrinsic spin-liquid physics in the kagome material herbertsmithite}},
	author = {Han, Tian-Heng and Norman, M. R. and Wen, J.-J. and Rodriguez-Rivera, Jose A. and Helton, Joel S. and Broholm, Collin and Lee, Young S.},
	journal = {Phys. Rev. B},
	volume = {94},
	issue = {6},
	pages = {060409},
	numpages = {5},
	year = {2016},
	month = {Aug},
	publisher = {American Physical Society},
	doi = {10.1103/PhysRevB.94.060409},
	url = {https://link.aps.org/doi/10.1103/PhysRevB.94.060409}
}

@ARTICLE{FuM15,
	author = "Mingxuan Fu and Takashi Imai and Tian-Heng Han and Young S. Lee",
	journal = "Science",
	volume = "350",
	pages = "655--658",
	year = "2015",
	title = {{Evidence for a gapped spin-liquid ground state in a kagome Heisenberg antiferromagnet}},
	doi = {10.1126/science.aab2120}
}

@ARTICLE{HanTH12,
	author = "T. H. Han and J. S. Helton and S. Chu and D. G. Nocera and J. A. Rodriguez-Rivera and C. Broholm and Y. S. Lee",
	journal = "Nature",
	volume = "492",
	pages = "406--410",
	year = "2012",
	title = {{Fractionalized excitations in the spin-liquid state of a kagome-lattice antiferromagnet}},
	url = {https://doi.org/10.1038/nature11659}
}

@ARTICLE{FengZL17,
	author = {Zili Feng and Zheng Li and Xin Meng and Wei Yi and Yuan Wei and Jun Zhang and Yan-Cheng Wang and Wei Jiang and Zheng Liu and Shiyan Li and Feng Liu and Jianlin Luo and Shiliang Li and Guo-qing Zheng and Zi Yang Meng and Jia-Wei Mei and Youguo Shi},
	title = {Gapped Spin-1/2 Spinon Excitations in a New Kagome Quantum Spin Liquid Compound {Cu$_3$Zn(OH)$_6$FBr}},
	year = {2017},
	journal = {Chin. Phys. Lett.},
	volume = {34},
	pages = {077502-077502},
	url = {http://cpl.iphy.ac.cn/en/article/doi/10.1088/0256-307X/34/7/077502},
	doi = {10.1088/0256-307X/34/7/077502}
}

@article{FengZL18b,
	title = {{From Claringbullite to a New Spin Liquid Candidate {Cu$_3$Zn(OH)$_6$FCl}}},
	author = "Zili Feng and Wei Yi and Kejia Zhu and Yuan Wei and Shanshan Miao and Jie Ma and Jianlin Luo and Shiliang Li and Zi Yang Meng and Youguo Shi",
	journal = "Chin. Phys. Lett.",
	year = "2018",
	volume = "36", 
	pages = "017502",
	doi = "10.1088/0256-307X/36/1/017502"
}

@article{HeYC17,
	title = {{Signatures of Dirac cones in a DMRG study of the Kagome Heisenberg model}},
	author = {He, Yin-Chen and Zaletel, Michael P. and Oshikawa, Masaki and Pollmann, Frank},
	journal = {Phys. Rev. X},
	volume = {7},
	issue = {3},
	pages = {031020},
	numpages = {16},
	year = {2017},
	month = {Jul},
	publisher = {American Physical Society},
	doi = {10.1103/PhysRevX.7.031020},
	url = {https://link.aps.org/doi/10.1103/PhysRevX.7.031020}
}

@article{HermeleM08,
	title = {{Properties of an algebraic spin liquid on the kagome lattice}},
	author = {Hermele, Michael and Ran, Ying and Lee, Patrick A. and Wen, Xiao-Gang},
	journal = {Phys. Rev. B},
	volume = {77},
	issue = {22},
	pages = {224413},
	numpages = {23},
	year = {2008},
	month = {Jun},
	publisher = {American Physical Society},
	doi = {10.1103/PhysRevB.77.224413},
	url = {https://link.aps.org/doi/10.1103/PhysRevB.77.224413}
}

@article{IqbalY14,
	title = {{Vanishing spin gap in a competing spin-liquid phase in the kagome Heisenberg antiferromagnet}},
	author = {Iqbal, Yasir and Poilblanc, Didier and Becca, Federico},
	journal = {Phys. Rev. B},
	volume = {89},
	issue = {2},
	pages = {020407},
	numpages = {5},
	year = {2014},
	month = {Jan},
	publisher = {American Physical Society},
	doi = {10.1103/PhysRevB.89.020407},
	url = {https://link.aps.org/doi/10.1103/PhysRevB.89.020407}
}

@article{DepenbrockS12,
	title = {{Nature of the Spin-Liquid Ground State of the $S=1/2$ {Heisenberg} Model on the Kagome Lattice}},
	author = {Depenbrock, Stefan and McCulloch, Ian P. and Schollw\"ock, Ulrich},
	journal = {Phys. Rev. Lett.},
	volume = {109},
	issue = {6},
	pages = {067201},
	numpages = {6},
	year = {2012},
	month = {Aug},
	publisher = {American Physical Society},
	doi = {10.1103/PhysRevLett.109.067201},
	url = {https://link.aps.org/doi/10.1103/PhysRevLett.109.067201}
}

@article{YanS11,
	title = {{Spin-Liquid Ground State of the S = 1/2 Kagome {Heisenberg} Antiferromagnet}},
	author = "Simeng Yan and David A. Huse and Steven R. White",
	journal = "Science",
	year = "2011",
	Volume =  "332", 
	pages = "1173",
	doi = "10.1126/science.1201080"
}

@article{BroholmC20,
	title = {{Quantum spin liquids}},
	author = "C. Broholm and R. J. Cava and S. A. Kivelson and D. G. Nocera and M. R. Norman and T. Senthil",
	journal = "Science",
	volume = "367",
	pages = "263",
	year = "2020",
	doi = "10.1126/science.aay0668"
}

@article{HastingsMB00,
	title = {{Dirac structure, RVB, and Goldstone modes in the kagom\'e antiferromagnet}},
	author = {Hastings, M. B.},
	journal = {Phys. Rev. B},
	volume = {63},
	issue = {1},
	pages = {014413},
	numpages = {16},
	year = {2000},
	month = {Dec},
	publisher = {American Physical Society},
	doi = {10.1103/PhysRevB.63.014413},
	url = {https://link.aps.org/doi/10.1103/PhysRevB.63.014413}
}

@article{RanY07,
	title = {{Projected-Wave-Function Study of the Spin-$1/2$ Heisenberg Model on the Kagom\'e Lattice}},
	author = {Ran, Ying and Hermele, Michael and Lee, Patrick A. and Wen, Xiao-Gang},
	journal = {Phys. Rev. Lett.},
	volume = {98},
	issue = {11},
	pages = {117205},
	numpages = {4},
	year = {2007},
	month = {Mar},
	publisher = {American Physical Society},
	doi = {10.1103/PhysRevLett.98.117205},
	url = {https://link.aps.org/doi/10.1103/PhysRevLett.98.117205}
}

@article{SavaryL17,
	title = {{Quantum spin liquids: {A} review}},
	author = {Lucile Savary and Leon Balents},
	journal = {Rep. Prog. Phys.},
	pages = {016502},
	url = {http://dx.doi.org/10.1088/0034-4885/80/1/016502},
	volume = {80},
	year = {2017}}

@article{BalentsL10,
	title = {{Spin liquids in frustrated magnets}},
	author = {Leon Balents},
	journal = {Nature},
	pages = {199},
	volume = {464},
	year = {2010},
	doi = "10.1038/nature08917"
}

@article{NormanMR16,
	author = {M. R. Norman},
	doi = {10.1103/RevModPhys.88.041002},
	journal = {Rev. Mod. Phys.},
	pages = {041002},
	title = {{Colloquium: Herbertsmithite and the search for the quantum spin liquid}},
	volume = {88},
	year = {2016}}

@ARTICLE{WeiYuan2017,
	author = {{Wei}, Yuan and {Feng}, Zili and {Lohstroh}, Wiebke and {Yu}, D.~H. and {Le}, Duc and {dela Cruz}, Clarina and {Yi}, Wei and {Ding}, Z.~F. and {Zhang}, J. and {Tan}, Cheng and {Shu}, Lei and {Wang}, Yan-Cheng and {Wu}, Han-Qing and {Luo}, Jianlin and {Mei}, Jia-Wei and {Yang}, Fang and {Sheng}, Xian-Lei and {Li}, Wei and {Qi}, Yang and {Meng}, Zi Yang and {Shi}, Youguo and {Li}, Shiliang},
	title = {{Evidence for the topological order in a kagome antiferromagnet}},
	journal = {arXiv e-prints},
	year = 2017,
	month = oct,
	eid = {arXiv:1710.02991},
	pages = {arXiv:1710.02991},
	archivePrefix = {arXiv},
	eprint = {1710.02991},
	primaryClass = {cond-mat.str-el},
	adsurl = {https://ui.adsabs.harvard.edu/abs/2017arXiv171002991W}
}

@article{Montecarlo2019Xu,
	title = {{Monte Carlo Study of Lattice Compact Quantum Electrodynamics with Fermionic Matter: The Parent State of Quantum Phases}},
	author = {Xu, Xiao Yan and Qi, Yang and Zhang, Long and Assaad, Fakher F. and Xu, Cenke and Meng, Zi Yang},
	journal = {Phys. Rev. X},
	volume = {9},
	issue = {2},
	pages = {021022},
	numpages = {17},
	year = {2019},
	month = {May},
	publisher = {American Physical Society},
	doi = {10.1103/PhysRevX.9.021022},
	url = {https://link.aps.org/doi/10.1103/PhysRevX.9.021022}
}

@ARTICLE{YYHuang2021,
	title = {{Heat Transport in Herbertsmithite: {Can} a Quantum Spin Liquid Survive Disorder?}},
	author = {Huang, Y. Y. and Xu, Y. and Wang, Le and Zhao, C. C. and Tu, C. P. and Ni, J. M. and Wang, L. S. and Pan, B. L. and Fu, Ying and Hao, Zhanyang and Liu, Cai and Mei, Jia-Wei and Li, S. Y.},
	journal = {Phys. Rev. Lett.},
	volume = {127},
	issue = {26},
	pages = {267202},
	numpages = {8},
	year = {2021},
	month = {Dec},
	publisher = {American Physical Society},
	doi = {10.1103/PhysRevLett.127.267202},
	url = {https://link.aps.org/doi/10.1103/PhysRevLett.127.267202}
}

@article{mengPerspective26,
author = {Meng, Zi Yang and Batista, Cristian D. and Li, Shiliang},
year = {2026},
title = {{An integrated theoretical and numerical approach to understand modern experiments on quantum magnetism}},
journal = {Nat. Phys.},
pages = {1189--1197},
volume = {22},
url = {https://doi.org/10.1038/s41567-026-03304-5},
doi = {10.1038/s41567-026-03304-5}
}

@article{dmrg_white_1992,
  title = {{Density matrix formulation for quantum renormalization groups}},
  author = {White, Steven R.},
  journal = {Phys. Rev. Lett.},
  volume = {69},
  issue = {19},
  pages = {2863--2866},
  numpages = {0},
  year = {1992},
  month = {Nov},
  publisher = {American Physical Society},
  doi = {10.1103/PhysRevLett.69.2863},
  url = {https://link.aps.org/doi/10.1103/PhysRevLett.69.2863}
}

@article{dmrg_white_1993,
  title = {{Density-matrix algorithms for quantum renormalization groups}},
  author = {White, Steven R.},
  journal = {Phys. Rev. B},
  volume = {48},
  issue = {14},
  pages = {10345--10356},
  numpages = {0},
  year = {1993},
  month = {Oct},
  publisher = {American Physical Society},
  doi = {10.1103/PhysRevB.48.10345},
  url = {https://link.aps.org/doi/10.1103/PhysRevB.48.10345}
}

@article{dmrg_ulrich_2011,
  title = {{The density-matrix renormalization group in the age of matrix product states}},
  author = {Schollw\"ock, U.},
  journal = {Annals of Physics},
  volume = {326},
  pages = {96-192},
  year = {2011},
  doi = {https://doi.org/10.1016/j.aop.2010.09.012},
  url = {https://www.sciencedirect.com/science/article/pii/S0003491610001752}
}

@article{dmrg_ulrich_2005,
  title = {{The density-matrix renormalization group}},
  author = {Schollw\"ock, U.},
  journal = {Rev. Mod. Phys.},
  volume = {77},
  issue = {1},
  pages = {259--315},
  numpages = {0},
  year = {2005},
  month = {Apr},
  publisher = {American Physical Society},
  doi = {10.1103/RevModPhys.77.259},
  url = {https://link.aps.org/doi/10.1103/RevModPhys.77.259}
}

@article{Amari1998,
    author = {Amari, Shun-ichi},
    title = {{Natural Gradient Works Efficiently in Learning}},
    journal = {Neural Computation},
    volume = {10},
    number = {2},
    pages = {251-276},
    year = {1998},
    month = {02},
    issn = {0899-7667},
    doi = {10.1162/089976698300017746},
    url = {https://doi.org/10.1162/089976698300017746},
}

@book{Becca_2017, address={Cambridge, England}, title={{Quantum Monte Carlo Approaches for Correlated Systems}}, publisher={Cambridge University Press}, author={Becca, Federico and Sorella, Sandro}, year={2017}}

@article{Chen2024,
author={Chen, Ao and Heyl, Markus},
title={{Empowering deep neural quantum states through efficient optimization}},
journal={Nat. Phys.},
year={2024},
month={Sep},
day={01},
volume={20},
number={9},
pages={1476-1481},
issn={1745-2481},
doi={10.1038/s41567-024-02566-1},
url={https://doi.org/10.1038/s41567-024-02566-1}
}

@article{Goldshlager2024,
title = {{A Kaczmarz-inspired approach to accelerate the optimization of neural network wavefunctions}},
journal = {Journal of Computational Physics},
volume = {516},
pages = {113351},
year = {2024},
issn = {0021-9991},
doi = {https://doi.org/10.1016/j.jcp.2024.113351},
url = {https://www.sciencedirect.com/science/article/pii/S0021999124005990},
author = {Gil Goldshlager and Nilin Abrahamsen and Lin Lin},
}

@article{Rende2024_potts,
  title = {{Mapping of attention mechanisms to a generalized Potts model}},
  author = {Rende, Riccardo and Gerace, Federica and Laio, Alessandro and Goldt, Sebastian},
  journal = {Phys. Rev. Res.},
  volume = {6},
  issue = {2},
  pages = {023057},
  numpages = {10},
  year = {2024},
  month = {Apr},
  publisher = {American Physical Society},
  doi = {10.1103/PhysRevResearch.6.023057},
  url = {https://link.aps.org/doi/10.1103/PhysRevResearch.6.023057}
}

@article{Rende2024,
author={Rende, Riccardo and Viteritti, Luciano Loris and Bardone, Lorenzo and Becca, Federico and Goldt, Sebastian},
title={{A simple linear algebra identity to optimize large-scale neural network quantum states}},
journal={Communications Physics},
year={2024},
month={Aug},
day={02},
volume={7},
number={1},
pages={260},
issn={2399-3650},
doi={10.1038/s42005-024-01732-4},
url={https://doi.org/10.1038/s42005-024-01732-4}
}

@article{Sorella1998,
  title = {Green Function {Monte} {Carlo} with Stochastic Reconfiguration},
  author = {Sorella, Sandro},
  journal = {Phys. Rev. Lett.},
  volume = {80},
  issue = {20},
  pages = {4558--4561},
  numpages = {0},
  year = {1998},
  month = {May},
  publisher = {American Physical Society},
  doi = {10.1103/PhysRevLett.80.4558},
  url = {https://link.aps.org/doi/10.1103/PhysRevLett.80.4558}
}

@article{Viteritti2023,
  title = {Transformer Variational Wave Functions for Frustrated Quantum Spin Systems},
  author = {Viteritti, Luciano Loris and Rende, Riccardo and Becca, Federico},
  journal = {Phys. Rev. Lett.},
  volume = {130},
  issue = {23},
  pages = {236401},
  numpages = {6},
  year = {2023},
  month = {Jun},
  publisher = {American Physical Society},
  doi = {10.1103/PhysRevLett.130.236401},
  url = {https://link.aps.org/doi/10.1103/PhysRevLett.130.236401}
}

@article{Viteritti2025,
  title = {{Transformer wave function for two dimensional frustrated magnets: Emergence of a spin-liquid phase in the Shastry-Sutherland model}},
  author = {Viteritti, Luciano Loris and Rende, Riccardo and Parola, Alberto and Goldt, Sebastian and Becca, Federico},
  journal = {Phys. Rev. B},
  volume = {111},
  issue = {13},
  pages = {134411},
  numpages = {15},
  year = {2025},
  month = {Apr},
  publisher = {American Physical Society},
  doi = {10.1103/PhysRevB.111.134411},
  url = {https://link.aps.org/doi/10.1103/PhysRevB.111.134411}
}

@article{Rende2025_QK,
  title = {Are Queries and Keys Always Relevant? {{A}} Case Study on Transformer Wave Functions},
  shorttitle = {Are Queries and Keys Always Relevant?},
  author = {Rende, Riccardo and Loris Viteritti, Luciano},
  year = {2025},
  month = mar,
  journal = {Machine Learning: Science and Technology},
  volume = {6},
  number = {1},
  pages = {010501},
  issn = {2632-2153},
  doi = {10.1088/2632-2153/ada1a0},
  urldate = {2025-08-27}
}

@article{JHC_PRL2008,
  title = {Density Matrix Renormalization Group Numerical Study of the Kagome Antiferromagnet},
  author = {Jiang, H. C. and Weng, Z. Y. and Sheng, D. N.},
  journal = {Phys. Rev. Lett.},
  volume = {101},
  issue = {11},
  pages = {117203},
  numpages = {4},
  year = {2008},
  month = {Sep},
  publisher = {American Physical Society},
  doi = {10.1103/PhysRevLett.101.117203},
  url = {https://link.aps.org/doi/10.1103/PhysRevLett.101.117203}
}

@article{JWM_PRB2017,
  title = {{Gapped spin liquid with ${\mathbb{Z}}_{2}$ topological order for the kagome Heisenberg model}},
  author = {Mei, Jia-Wei and Chen, Ji-Yao and He, Huan and Wen, Xiao-Gang},
  journal = {Phys. Rev. B},
  volume = {95},
  issue = {23},
  pages = {235107},
  numpages = {9},
  year = {2017},
  month = {Jun},
  publisher = {American Physical Society},
  doi = {10.1103/PhysRevB.95.235107},
  url = {https://link.aps.org/doi/10.1103/PhysRevB.95.235107}
}

@article{JHC_NP2012,
  title = {{Identifying topological order by entanglement entropy}},
  author = {Jiang, Hong-Chen and Wang, Zhenghan and Balents, Leon},
  journal = {Nat. Phys.},
  volume = {8},
  pages = {902--905},
  year = {2012},
  doi = {10.1038/nphys2465},
  url = {https://doi.org/10.1038/nphys2465}
}

@article{GSS_SR2014,
  title = {Emergent Chiral Spin Liquid: Fractional Quantum {Hall} Effect in a Kagome {Heisenberg} Model},
  author = {Gong, Shou-Shu and Zhu, Wei and Sheng, D. N.},
  journal = {Scientific Reports},
  volume = {4},
  pages = {6317},
  year = {2014},
  doi = {10.1038/srep06317},
  url = {https://doi.org/10.1038/srep06317}
}

@ARTICLE{li2016z2spinliquidphase,
	author = {Tao Li},
	title = {{Z$_{2}$ spin liquid phase on the kagome lattice: a new saddle point}},
	journal = {arXiv e-prints},
	year = {2016},
	archivePrefix = {arXiv},
	eprint = {1601.02165},
	primaryClass = {cond-mat.str-el},
    url={https://arxiv.org/abs/1601.02165}, 
}

@article{SRY_npjQM2024,
  title = {{Possible chiral spin liquid state in the S = 1/2 kagome Heisenberg model}},
  author = {Sun, Rong-Yang and Jin, Hui-Ke and Tu, Hong-Hao and Zhou, Yi},
  journal = {npj Quantum Materials},
  volume = {9},
  pages = {16},
  year = {2024},
  doi = {10.1038/s41535-024-00627-5},
  url = {https://doi.org/10.1038/s41535-024-00627-5}
}

@article{Messio_PRL2012,
  title = {Kagome Antiferromagnet: A Chiral Topological Spin Liquid?},
  author = {Messio, Laura and Bernu, Bernard and Lhuillier, Claire},
  journal = {Phys. Rev. Lett.},
  volume = {108},
  issue = {20},
  pages = {207204},
  numpages = {5},
  year = {2012},
  month = {May},
  publisher = {American Physical Society},
  doi = {10.1103/PhysRevLett.108.207204},
  url = {https://link.aps.org/doi/10.1103/PhysRevLett.108.207204}
}

@article{Iqbal_PRB2011,
  title = {{Valence-bond crystal in the extended kagome spin-$\frac{1}{2}$ quantum Heisenberg antiferromagnet: A variational Monte Carlo approach}},
  author = {Iqbal, Yasir and Becca, Federico and Poilblanc, Didier},
  journal = {Phys. Rev. B},
  volume = {83},
  issue = {10},
  pages = {100404(R)},
  numpages = {4},
  year = {2011},
  month = {Mar},
  publisher = {American Physical Society},
  doi = {10.1103/PhysRevB.83.100404},
  url = {https://link.aps.org/doi/10.1103/PhysRevB.83.100404}
}

@article{LHJXT_PRL2017,
  title = {Gapless Spin-Liquid Ground State in the {$S=1/2$} Kagome Antiferromagnet},
  author = {Liao, H. J. and Xie, Z. Y. and Chen, J. and Liu, Z. Y. and Xie, H. D. and Huang, R. Z. and Normand, B. and Xiang, T.},
  journal = {Phys. Rev. Lett.},
  volume = {118},
  issue = {13},
  pages = {137202},
  numpages = {6},
  year = {2017},
  month = {Mar},
  publisher = {American Physical Society},
  doi = {10.1103/PhysRevLett.118.137202},
  url = {https://link.aps.org/doi/10.1103/PhysRevLett.118.137202}
}

@article{liuGapless2022,
  title = {Gapless spin liquid behavior in a kagome Heisenberg antiferromagnet with randomly distributed hexagons of alternate bonds},
  author = {Liu, Jiabin and Yuan, Long and Li, Xuan and Li, Boqiang and Zhao, Kan and Liao, Haijun and Li, Yuesheng},
  journal = {Phys. Rev. B},
  volume = {105},
  issue = {2},
  pages = {024418},
  numpages = {20},
  year = {2022},
  month = {Jan},
  publisher = {American Physical Society},
  doi = {10.1103/PhysRevB.105.024418},
  url = {https://link.aps.org/doi/10.1103/PhysRevB.105.024418}
}

@article{IqbalBecca_PRB2013,
  title = {{Gapless spin-liquid phase in the kagome spin-$\frac{1}{2}$ Heisenberg antiferromagnet}},
  author = {Iqbal, Yasir and Becca, Federico and Sorella, Sandro and Poilblanc, Didier},
  journal = {Phys. Rev. B},
  volume = {87},
  issue = {6},
  pages = {060405(R)},
  numpages = {5},
  year = {2013},
  month = {Feb},
  publisher = {American Physical Society},
  doi = {10.1103/PhysRevB.87.060405},
  url = {https://link.aps.org/doi/10.1103/PhysRevB.87.060405}
}

@article{Hu_PRB2015,
  title = {{Variational Monte Carlo study of a gapless spin liquid in the spin-$\frac{1}{2}$ XXZ antiferromagnetic model on the kagome lattice}},
  author = {Hu, Wen-Jun and Gong, Shou-Shu and Becca, Federico and Sheng, D. N.},
  journal = {Phys. Rev. B},
  volume = {92},
  issue = {20},
  pages = {201105(R)},
  numpages = {5},
  year = {2015},
  month = {Nov},
  publisher = {American Physical Society},
  doi = {10.1103/PhysRevB.92.201105},
  url = {https://link.aps.org/doi/10.1103/PhysRevB.92.201105}
}

@article{JiangSH_2019SciPostPhys,
	title = {{Competing Spin Liquid Phases in the S=$\frac{1}{2}$ Heisenberg Model on the Kagome Lattice}},
	pages = {006},
	author = {Jiang, Shenghan and Kim, Panjin and Han, Jung Hoon and Ran, Ying},
	journal = {SciPost Phys.},
	volume = {7},
	year = {2019},
	publisher = {SciPost},
	doi = {10.21468/SciPostPhys.7.1.006},
	url = {https://scipost.org/10.21468/SciPostPhys.7.1.006}
}

@article{SinghVBC_PRB2007,
  title = {{Ground state of the spin-1/2 kagome-lattice Heisenberg antiferromagnet}},
  author = {Singh, Rajiv R. P. and Huse, David A.},
  journal = {Phys. Rev. B},
  volume = {76},
  issue = {18},
  pages = {180407R},
  numpages = {4},
  year = {2007},
  month = {Nov},
  publisher = {American Physical Society},
  doi = {10.1103/PhysRevB.76.180407},
  url = {https://link.aps.org/doi/10.1103/PhysRevB.76.180407}
}

@article{Budnik_PRL2004,
  title = {Low-Energy Singlets in the {Heisenberg} Antiferromagnet on the Kagome Lattice},
  author = {Budnik, Ran and Auerbach, Assa},
  journal = {Phys. Rev. Lett.},
  volume = {93},
  issue = {18},
  pages = {187205},
  numpages = {4},
  year = {2004},
  month = {Oct},
  publisher = {American Physical Society},
  doi = {10.1103/PhysRevLett.93.187205},
  url = {https://link.aps.org/doi/10.1103/PhysRevLett.93.187205}
}

@article{Evenbly_PRL2020,
  title = {{Frustrated Antiferromagnets with Entanglement Renormalization: Ground State of the Spin-$\frac{1}{2}$ Heisenberg Model on a Kagome Lattice}},
  author = {Evenbly, G. and Vidal, G.},
  journal = {Phys. Rev. Lett.},
  volume = {104},
  issue = {18},
  pages = {187203},
  numpages = {4},
  year = {2010},
  month = {May},
  publisher = {American Physical Society},
  doi = {10.1103/PhysRevLett.104.187203},
  url = {https://link.aps.org/doi/10.1103/PhysRevLett.104.187203}
}

@article{ZWGSSDNS_2025QF,
  title = {{Quantum spin liquids in frustrated Kagome Heisenberg model}},
  author = {Zhu, W. and Gong, Shou-Shu and Sheng, D. N.},
  journal = {Quantum Frontiers},
  volume = {4},
  pages = {11},
  year = {2025},
  doi = {10.1007/s44214-025-00084-6},
  url = {https://doi.org/10.1007/s44214-025-00084-6}
}

@article{Carleo2017,
author = {Giuseppe Carleo  and Matthias Troyer },
title = {{Solving the quantum many-body problem with artificial neural networks}},
journal = {Science},
volume = {355},
number = {6325},
pages = {602-606},
year = {2017},
doi = {10.1126/science.aag2302},
URL = {https://www.science.org/doi/abs/10.1126/science.aag2302},
}

@article{Lange_2024,
doi = {10.1088/2058-9565/ad7168},
url = {https://doi.org/10.1088/2058-9565/ad7168},
year = {2024},
month = {sep},
publisher = {IOP Publishing},
volume = {9},
number = {4},
pages = {040501},
author = {Lange, Hannah and Van de Walle, Anka and Abedinnia, Atiye and Bohrdt, Annabelle},
title = {{From architectures to applications: a review of neural quantum states}},
journal = {Quantum Science and Technology},
}

@article{hpc2021,
journal={HPC2021, Information Technology Services, The University of Hong Kong},
url={https://hpc.hku.hk/hpc/hpc2021/}
}

@ARTICLE{paratera,
journal={Beijing PARATERA
Tech CO.,Ltd},
url = {https://cloud.paratera.com}
}

@article{knolleField2019,
   author = "Knolle, J. and Moessner, R.",
   title = "A Field Guide to Spin Liquids", 
   journal= "Annual Review of Condensed Matter Physics",
   year = "2019",
   volume = "10",
   number = "Volume 10, 2019",
   pages = "451-472",
   doi = "https://doi.org/10.1146/annurev-conmatphys-031218-013401",
   url = "https://www.annualreviews.org/content/journals/10.1146/annurev-conmatphys-031218-013401",
   publisher = "Annual Reviews",
   issn = "1947-5462",
   type = "Journal Article"
  }

@article{Hu_DSL_PRL2019,
  title = {{Dirac Spin Liquid on the Spin-$1/2$ Triangular Heisenberg Antiferromagnet}},
  author = {Hu, Shijie and Zhu, W. and Eggert, Sebastian and He, Yin-Chen},
  journal = {Phys. Rev. Lett.},
  volume = {123},
  issue = {20},
  pages = {207203},
  numpages = {6},
  year = {2019},
  month = {Nov},
  publisher = {American Physical Society},
  doi = {10.1103/PhysRevLett.123.207203},
  url = {https://link.aps.org/doi/10.1103/PhysRevLett.123.207203}
}

@article{Jiang_DDMRG_2026,
  title = {{Competing States in the $S=1/2$ Triangular-Lattice ${J}_{1}\text{\ensuremath{-}}{J}_{2}$ Heisenberg Model: A Dynamical Density-Matrix Renormalization Group Study}},
  author = {Jiang, Shengtao and White, Steven R. and Kivelson, Steven A. and Jiang, Hong-Chen},
  journal = {Phys. Rev. Lett.},
  volume = {137},
  issue = {5},
  pages = {056703},
  numpages = {10},
  year = {2026},
  month = {Jul},
  publisher = {American Physical Society},
  doi = {10.1103/zmnz-tkq2},
  url = {https://link.aps.org/doi/10.1103/zmnz-tkq2}
}

@article{Xiang_PRL2026,
  title = {Dynamical Spectral Function of the Kagome Quantum Spin Liquid},
  author = {Hu, Jiahang and Chi, Runze and Guo, Yibin and Normand, B. and Liao, Hai-Jun and Xiang, T.},
  journal = {Phys. Rev. Lett.},
  volume = {137},
  issue = {7},
  pages = {076504},
  numpages = {8},
  year = {2026},
  month = {Aug},
  publisher = {American Physical Society},
  doi = {10.1103/fjjf-xspp},
  url = {https://link.aps.org/doi/10.1103/fjjf-xspp}
}

@article{Pollmann_PRB2023,
  title = {{Dynamical signatures of symmetry-broken and liquid phases in an $S$ = $\frac{1}{2}$ Heisenberg antiferromagnet on the triangular lattice}},
  author = {Drescher, Markus and Vanderstraeten, Laurens and Moessner, Roderich and Pollmann, Frank},
  journal = {Phys. Rev. B},
  volume = {108},
  issue = {22},
  pages = {L220401},
  numpages = {6},
  year = {2023},
  month = {Dec},
  publisher = {American Physical Society},
  doi = {10.1103/PhysRevB.108.L220401},
  url = {https://link.aps.org/doi/10.1103/PhysRevB.108.L220401}
}

@article{Bader_PRB2026,
  title = {{Triangular ${J}_{1}\text{\ensuremath{-}}{J}_{2}$ Heisenberg antiferromagnet in a magnetic field}},
  author = {Bader, T. and Feng, S. and Budaraju, S. and Becca, F. and Knolle, J. and Pollmann, F.},
  journal = {Phys. Rev. B},
  volume = {113},
  issue = {24},
  pages = {L241114},
  numpages = {7},
  year = {2026},
  month = {Jun},
  publisher = {American Physical Society},
  doi = {10.1103/rwhz-jf5b},
  url = {https://link.aps.org/doi/10.1103/rwhz-jf5b}
}

@misc{Starykh2025,
  title={$\mathrm{J}_1 - \mathrm{J}_2$ Triangular Lattice Antiferromagnet in a Magnetic Field},
  author={Anna Keselman and Xinyuan Xu and Hao Zhang and Cristian D. Batista and Oleg A. Starykh},
  year={2025},
  eprint={2512.02150},
  archivePrefix={arXiv},
  primaryClass={cond-mat.str-el},
  url={https://arxiv.org/abs/2512.02150},
  note={arXiv preprint}
}

@article{song2026ss,
  title={Thermodynamics of Shastry-Sutherland Model under Magnetic Field},
  author={Menghan Song and Chengkang Zhou and Cheng Huang and Zi Yang Meng},
  journal={arXiv preprint arXiv:2602.11589},
  year={2026},
  url={https://arxiv.org/abs/2602.11589}
}

@article{Roth_PRB2023,
  title = {{High-accuracy variational Monte Carlo for frustrated magnets with deep neural networks}},
  author = {Roth, Christopher and Szab\'o, Attila and MacDonald, Allan H.},
  journal = {Phys. Rev. B},
  volume = {108},
  issue = {5},
  pages = {054410},
  numpages = {12},
  year = {2023},
  month = {Aug},
  publisher = {American Physical Society},
  doi = {10.1103/PhysRevB.108.054410},
  url = {https://link.aps.org/doi/10.1103/PhysRevB.108.054410}
}

@article{Guoxin_PRX2025,
  title = {Thermodynamic Evidence of Fermionic Behavior in the Vicinity of One-Ninth Plateau in a {Kagome} Antiferromagnet},
  author = {Zheng, Guoxin and Zhang, Dechen and Zhu, Yuan and Chen, Kuan-Wen and Chan, Aaron and Jenkins, Kaila and Kang, Byungmin and Zeng, Zhenyuan and Xu, Aini and Ratkovski, D. and Blawat, Joanna and Bangura, Alimamy F. and Singleton, John and Lee, Patrick A. and Li, Shiliang and Li, Lu},
  journal = {Phys. Rev. X},
  volume = {15},
  issue = {2},
  pages = {021076},
  numpages = {15},
  year = {2025},
  month = {May},
  publisher = {American Physical Society},
  doi = {10.1103/PhysRevX.15.021076},
  url = {https://link.aps.org/doi/10.1103/PhysRevX.15.021076}
}

@article{Kumar_PRL2026,
  title = {{Dirac Node Pinning from Dzyaloshinskii-Moriya Interactions in a Kagome Spin Liquid}},
  author = {Kumar, Ajesh and Kang, Byungmin and Lee, Patrick A.},
  journal = {Phys. Rev. Lett.},
  volume = {136},
  issue = {14},
  pages = {146704},
  numpages = {7},
  year = {2026},
  month = {Apr},
  publisher = {American Physical Society},
  doi = {10.1103/wz27-6nhx},
  url = {https://link.aps.org/doi/10.1103/wz27-6nhx}
}

@misc{Byungmin_2025,
  title={Cyclotron resonance in a kagome spin liquid candidate material},
  author={Byungmin Kang and Patrick A. Lee},
  year={2025},
  eprint={2507.19576},
  archivePrefix={arXiv},
  primaryClass={cond-mat.str-el},
  url={https://arxiv.org/abs/2507.19576},
  note={arXiv preprint}
}

@misc{Rigo_2026,
      title={Neural quantum states in condensed matter: advances, best practices, and prospects}, 
      author={Jonas B. Rigo and Björn J. Wurst and Rajah Nutakki and Markus Schmitt and Dante Kennes},
      year={2026},
      eprint={2608.21291},
      archivePrefix={arXiv},
      primaryClass={cond-mat.str-el},
      url={https://arxiv.org/abs/2608.21291}, 
}

@software{hessel2020optax,
  title = {The {D}eep{M}ind {JAX} {E}cosystem},
  author = {DeepMind and Babuschkin, Igor and Baumli, Kate and Bell, Alison and Bhupatiraju, Surya and Bruce, Jake and Buchlovsky, Peter and Budden, David and Cai, Trevor and Clark, Aidan and Danihelka, Ivo and Dedieu, Antoine and Fantacci, Claudio and Godwin, Jonathan and Jones, Chris and Hemsley, Ross and Hennigan, Tom and Hessel, Matteo and Hou, Shaobo and Kapturowski, Steven and Keck, Thomas and Kemaev, Iurii and King, Michael and Kunesch, Markus and Martens, Lena and Merzic, Hamza and Mikulik, Vladimir and Norman, Tamara and Papamakarios, George and Quan, John and Ring, Roman and Ruiz, Francisco and Sanchez, Alvaro and Sartran, Laurent and Schneider, Rosalia and Sezener, Eren and Spencer, Stephen and Srinivasan, Srivatsan and Stanojevi\'{c}, Milo\v{s} and Stokowiec, Wojciech and Wang, Luyu and Zhou, Guangyao and Viola, Fabio},
  url = {http://github.com/google-deepmind},
  year = {2020},
}

@misc{Loschilov,
      title={SGDR: Stochastic Gradient Descent with Warm Restarts}, 
      author={Ilya Loshchilov and Frank Hutter},
      year={2017},
      eprint={1608.03983},
      archivePrefix={arXiv},
      primaryClass={cs.LG},
      url={https://arxiv.org/abs/1608.03983}, 
}

@ARTICLE{guoTlinear2026,
       author = {{Guo}, Jing and {Wang}, Pengyu and {Huang}, Cheng and {Zhou}, Chengkang and {Song}, Menghan and {Chen}, Xintian and {Wang}, Ting-Tung and {Hong}, Wenshan and {Cai}, Shu and {Zhao}, Jinyu and {Han}, Jinyu and {Zhou}, Yazhou and {Wu}, Qi and {Li}, Shiliang and {Meng}, Zi Yang and {Sun}, Liling},
        title = "{T-linear specific heat in pressurized and magnetized Shastry-Sutherland Mott insulator SrCu$_2$(BO$_3$)$_2$}",
      journal = {arXiv e-prints},
         year = 2026,
        month = feb,
          eid = {arXiv:2602.18229},
        pages = {arXiv:2602.18229},
          doi = {10.48550/arXiv.2602.18229},
archivePrefix = {arXiv},
       eprint = {2602.18229},
 primaryClass = {cond-mat.str-el},
       adsurl = {https://ui.adsabs.harvard.edu/abs/2026arXiv260218229G}
}

@ARTICLE{raikosVariational2026,
  title = {{Variational study of the magnetization plateaus in the spin-$\frac{1}{2}$ kagome Heisenberg antiferromagnet: An approach from vision transformer neural quantum states}},
  author = {Raikos, Andreas and Capponi, Sylvain and Alet, Fabien},
  journal = {Phys. Rev. B},
  volume = {114},
  issue = {7},
  pages = {074410},
  numpages = {16},
  year = {2026},
  month = {Aug},
  publisher = {American Physical Society},
  doi = {10.1103/xyzw-jtn1},
  url = {https://link.aps.org/doi/10.1103/xyzw-jtn1}
}

@misc{Viteritti2026thermolimit,
      title={Approaching the Thermodynamic Limit with Neural-Network Quantum States},
      author={Luciano Loris Viteritti and Riccardo Rende and Subir Sachdev and Giuseppe Carleo},
      year={2026},
      eprint={2602.02665},
      archivePrefix={arXiv},
      url={https://arxiv.org/abs/2602.02665},
}

@article{Capponi2013,
  title = {{$p6$ chiral resonating valence bonds in the kagome antiferromagnet}},
  author = {Capponi, Sylvain and Chandra, V. Ravi and Auerbach, Assa and Weinstein, Marvin},
  journal = {Phys. Rev. B},
  volume = {87},
  issue = {16},
  pages = {161118(R)},
  numpages = {5},
  year = {2013},
  month = {Apr},
  publisher = {American Physical Society},
  doi = {10.1103/PhysRevB.87.161118},
  url = {https://link.aps.org/doi/10.1103/PhysRevB.87.161118}
}

@misc{Rende2026scaling,
      title={Scaling Laws for Neural-Network Quantum States}, 
      author={Riccardo Rende and Alessandro Sinibaldi and Luciano Loris Viteritti and Roeland Wiersema and Antoine Georges and Giuseppe Carleo},
      year={2026},
      eprint={2606.02794},
      archivePrefix={arXiv},
      primaryClass={cond-mat.dis-nn},
      url={https://arxiv.org/abs/2606.02794}, 
}

@article{Nomura2021,
  title = {Dirac-Type Nodal Spin Liquid Revealed by Refined Quantum Many-Body Solver Using Neural-Network Wave Function, Correlation Ratio, and Level Spectroscopy},
  author = {Nomura, Yusuke and Imada, Masatoshi},
  journal = {Phys. Rev. X},
  volume = {11},
  issue = {3},
  pages = {031034},
  numpages = {19},
  year = {2021},
  month = {Aug},
  publisher = {American Physical Society},
  doi = {10.1103/PhysRevX.11.031034},
  url = {https://link.aps.org/doi/10.1103/PhysRevX.11.031034}
}

@article{Nomura2021_thermal,
  title = {Purifying Deep Boltzmann Machines for Thermal Quantum States},
  author = {Nomura, Yusuke and Yoshioka, Nobuyuki and Nori, Franco},
  journal = {Phys. Rev. Lett.},
  volume = {127},
  issue = {6},
  pages = {060601},
  numpages = {7},
  year = {2021},
  month = {Aug},
  publisher = {American Physical Society},
  doi = {10.1103/PhysRevLett.127.060601},
  url = {https://link.aps.org/doi/10.1103/PhysRevLett.127.060601}
}

@misc{TensorKit,
      title={{TensorKit.jl: A Julia package for large-scale tensor computations, with a hint of category theory}}, 
      author={Lukas Devos and Jutho Haegeman},
      year={2026},
      eprint={2508.10076},
      archivePrefix={arXiv},
      url={https://arxiv.org/abs/2508.10076}, 
}

\newpage
\clearpage
\onecolumngrid
\begin{center}
	\textbf{\large Supplemental Materials for: “Unraveling the Kagome Antiferromagnetic $3J$ Model and Its Materials: An Integrated Approach”}
\end{center}
\setcounter{equation}{0}
\setcounter{figure}{0}
\setcounter{table}{0}
\setcounter{page}{1}
\setcounter{section}{0}

\makeatletter
\renewcommand{\theequation}{S\arabic{equation}}
\renewcommand{\thefigure}{S\arabic{figure}}
\setcounter{secnumdepth}{3}
\setcounter{page}{1}
\setcounter{equation}{0}
\setcounter{figure}{0}
\renewcommand{\theequation}{S\arabic{equation}}
\renewcommand{\thefigure}{S\arabic{figure}}
\setcounter{secnumdepth}{3}

In the Supplemental Material, we provide additional results and methodological details supporting the conclusions presented in the main text. 
In Sec.~I, we present additional ground-state results from DMRG and NQS calculations, including the evolution of the static spin structure factor, real-space spin correlations, and bond energies across the phase diagram.
In Sec.~II, we describe the algorithmic framework and implementation details of the NQS approach used to obtain the ground states.
In Sec.~III, we provide further technical details of the XTRG and tanTRG thermal tensor-network algorithms, together with the specific heat results for additional parameter points.

%%%%%%%%%%%%%%%%%%%%%%%%%%%%%%%%%%%%%%%%%%%%%%%%%%%%%%%%%%%%%%%%%%%%%%%
\section{\label{sec:GS}Additional Ground-State Results by DMRG and NQS}

\begin{figure*}[h]
   \includegraphics[width=1.0\textwidth,angle=0]{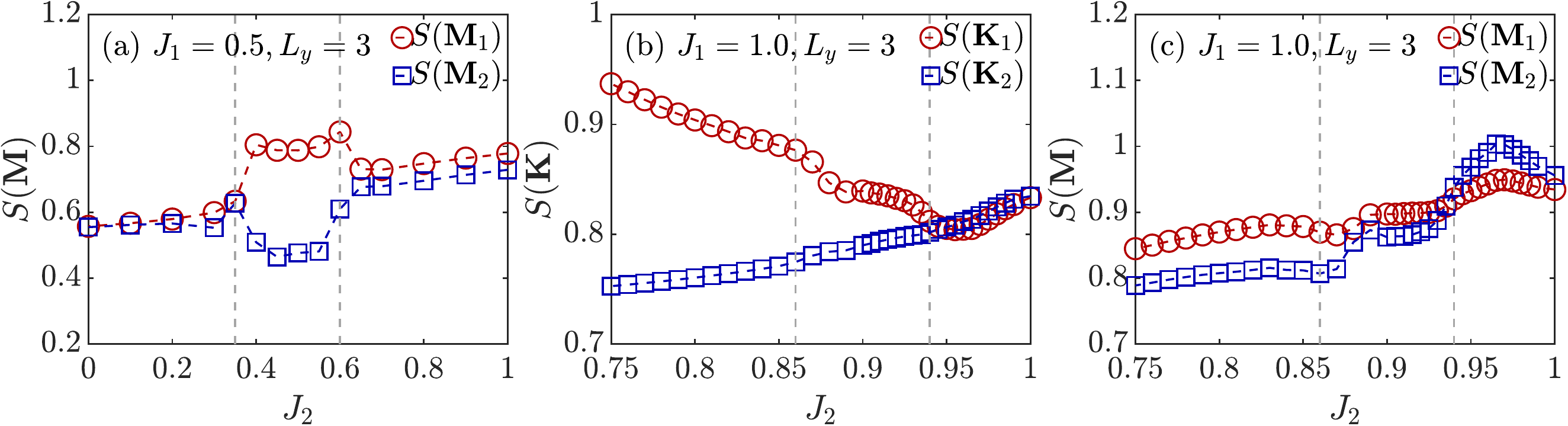}
   \caption{
   \label{PTDMRG}
  \textbf{$J_2$ dependence of $S(\mathbf{k})$ at the $\mathbf{M}$ and $\mathbf{K}$ points from DMRG calculations.} (a) $J_2$ dependence of $S(\mathbf{k})$ at $\mathbf{M}_1$ and $\mathbf{M}_2$ for $J_1=0.5$ on YC6 ($L_y=3$, $L_x=12$) cylinders. (b,c) Same as (a), but for $J_1=1.0$, with $S(\mathbf{k})$ evaluated at the $\mathbf{K}$ and $\mathbf{M}$ points, respectively. The kink-like features are indicated by the gray dashed lines.
   }   
\end{figure*}

In the main text, we identify the phase transition points by tracking kink-like features in the evolution of $S(\mathbf{K})$ in the $(1/3,1/3)$ ordered phases and $S(\mathbf{M})$ in the QSL and KSL phases, together with singular behavior in the entanglement entropy. Here, we provide an additional diagnostic based on the static spin structure factor, using $S(\mathbf{M})$ in the $(1/3,1/3)$ ordered phases and $S(\mathbf{K})$ in the QSL and KSL phases to independently cross-check the transition points. In Fig.~\ref{PTDMRG}(a), we show the evolution of $S(\mathbf{M}_1)$ and $S(\mathbf{M}_2)$ as a function of $J_2$ for $J_1=0.5$, obtained from DMRG calculations. Two pronounced kink-like features are observed near $J_2\simeq0.35$ and $0.6$, respectively, signaling changes in the underlying spin correlations and providing independent support for the phase boundaries shown in Figs.~\ref{PhaseTransition}(a-c). For $J_1=1.0$, similar kink-like features emerge near $J_2\simeq0.86$ and $0.94$ [cf. Fig.~\ref{PTDMRG}(b)]. The first kink, near $J_2\simeq0.86$, marks the transition from the $(1/3,1/3)$ ordered phase to the QSL phase and is also observed in the evolution of $S(\mathbf{M})$ [cf. Fig.~\ref{PTDMRG}(c)]. The second kink, near $J_2\simeq0.94$, agrees excellently with the QSL–KSL transition point identified from the complementary diagnostics in Figs.~\ref{PhaseTransition}(d-f), further supporting the robustness of the phase boundaries established from our ground-state calculations.

% \begin{figure*}[h]
%     \includegraphics[width=0.65\textwidth,angle=0]{FigSM_SM12NQS.pdf}
%     \caption{
%     \label{PTDMRG_NQS}
%    \textbf{$J_2$ dependence of $S(\mathbf{k})$ at the $\mathbf{M}$ and $\mathbf{K}$ points from NQS calculations.} (a) $J_2$ dependence of $S(\mathbf{k})$ at ${\mathbf{M}}_1$ and ${\mathbf{M}}_2$ for $J_1=0.5$ on a torus ($L_y=6, L_x=6$). (b) Same as (a), but for $J_1=1.0$, with $S(\mathbf{k})$ evaluated at ${\mathbf{K}}_1$ and ${\mathbf{K}}_2$. The kink-like features are indicated by the gray dashed lines.
%     }   
%  \end{figure*}

\begin{figure*}[h]
   \includegraphics[width=0.7\textwidth]{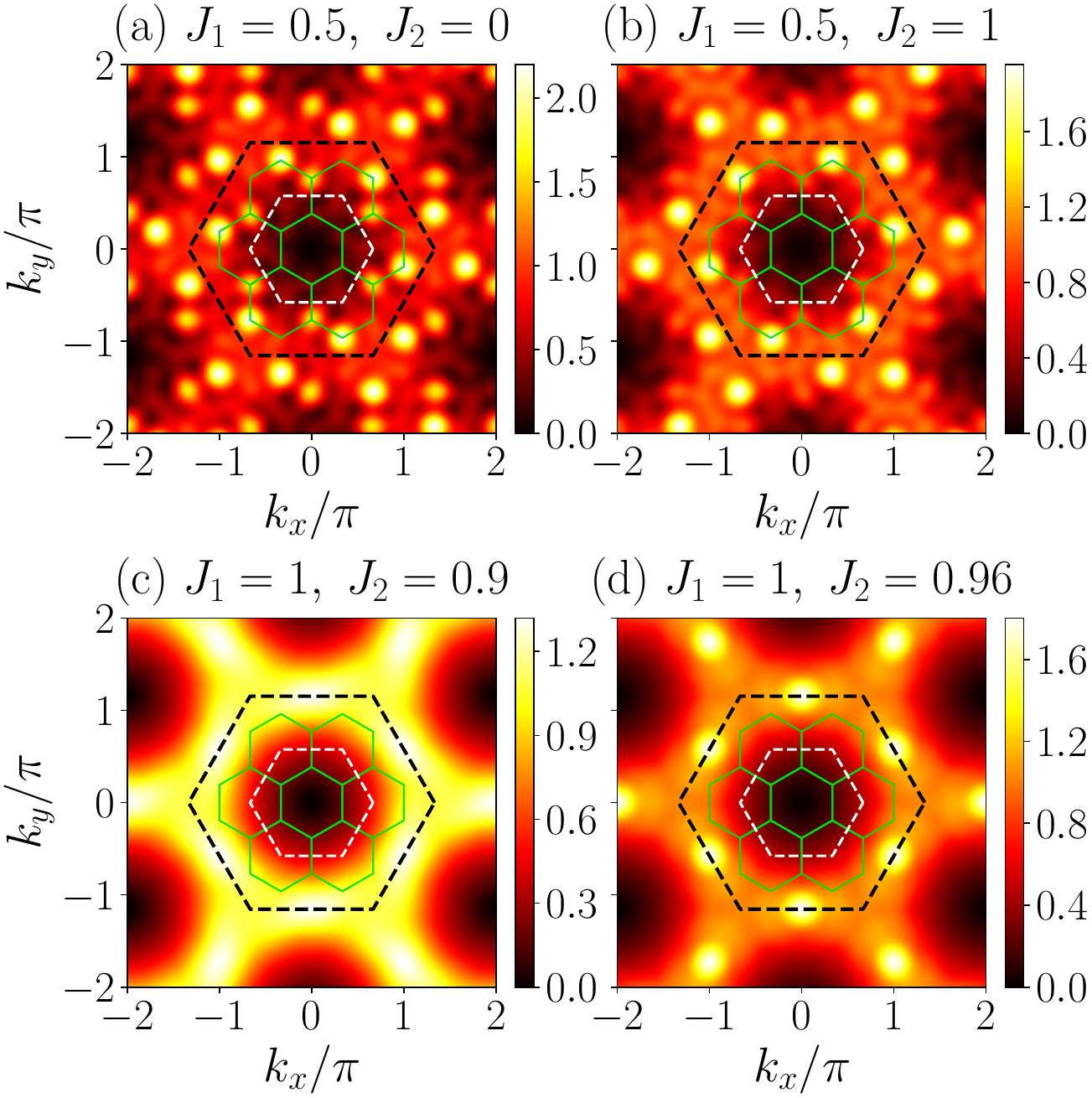}
   \caption{\textbf{Static spin structure factors from NQS.} Representative static spin structure factors $S(\mathbf{k})$ obtained from the NQS calculations on the $6\times6\times3$ torus in the (a,b) $(1/3,1/3)$ ordered phases, (c) QSL phase, and (d) KSL phase, with couplings in units of $J_{\hexagon}$. Pronounced peaks occur in (a,b) at the ordering vectors $\mathbf{Q}=\mathbf{K}_1$ and $\mathbf{Q}=\mathbf{K}_2$, respectively, at the same positions as in Figs.~\ref{SK}(a,b). Brillouin zone conventions are as in Fig.~\ref{SK}. All maps are Fourier transforms of the all-to-all spin-spin correlations  on the torus, estimated with $2^{16}$ Monte Carlo samples and plotted as continuous functions of $\mathbf{k}$.
   \label{SkNQS}
   }
\end{figure*}

\begin{figure*}[h]
   \includegraphics[width=0.95\textwidth,angle=0]{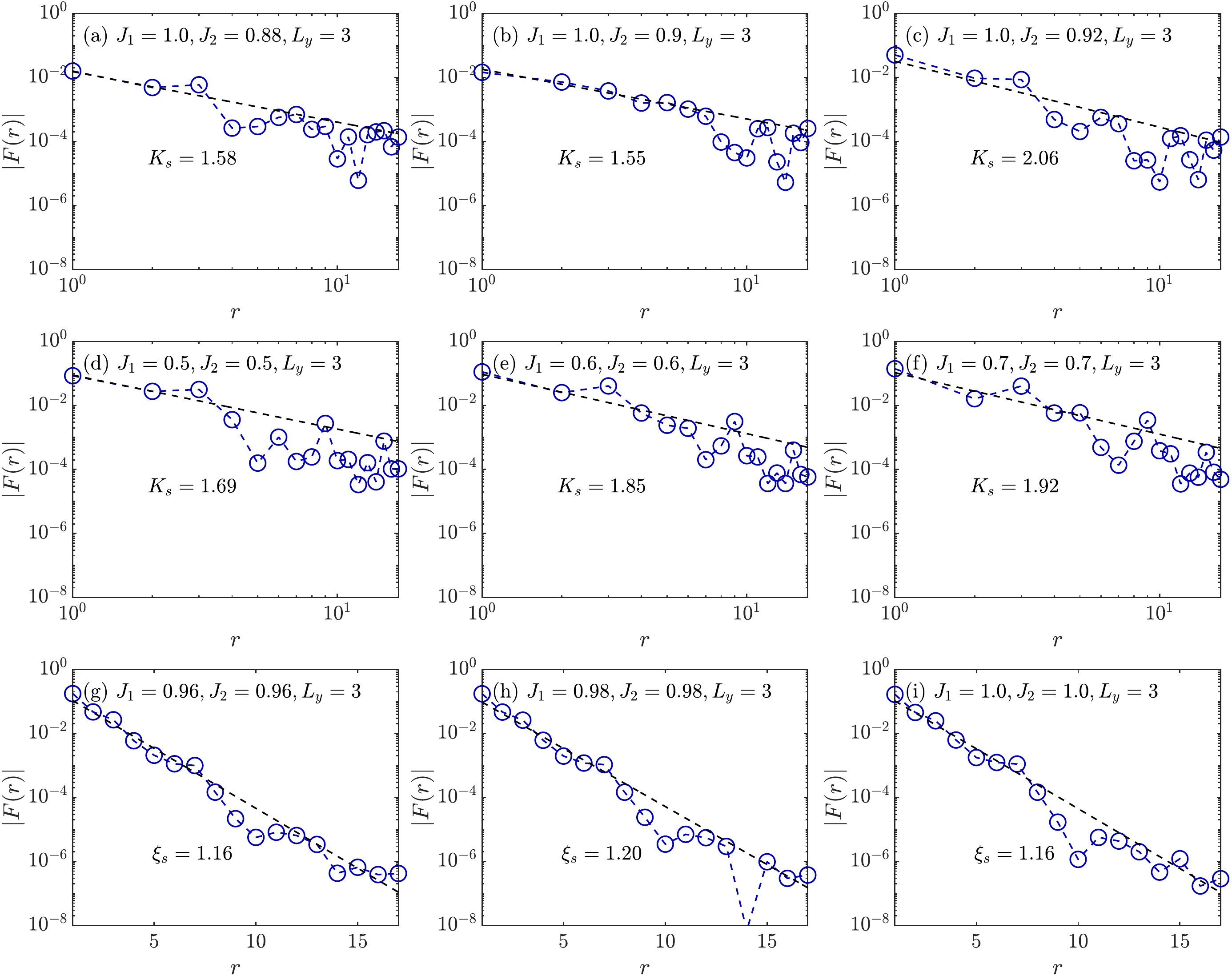}
   \caption{
   \label{SMSpin}
  \textbf{Spin correlations in the QSL and KSL phases.} (a-f) Double-logarithmic plots of the real-space spin correlation function $F(r)$ in the QSL phase. The power exponents $K_{\mathrm{s}}$ are obtained by algebraic fitting with dash line. (g-i) Semi-logarithmic plots of $F(r)$ in the KSL phase. The correlation lengths $\xi_{\mathrm{s}}$ are extracted by exponential fitting with dash line. All results are calculated on YC6 cylinders with $L_x=12$, using a maximum bond dimension of $D=15000$.
   }   
\end{figure*}

In Fig.~\ref{SkNQS}, we also present the $S(\mathbf{k})$ profiles obtained from NQS calculations, which are in excellent agreement with those shown in Fig.~\ref{SK}. One difference is that, in the QSL phase, $S(\mathbf{k})$ remains continuous at $\mathbf{M}_2$, in contrast to the behavior observed in Fig.~\ref{SK}(c). This difference can be attributed to the distinct cylinder and torus geometries, with the latter restoring the $C_3$ rotational symmetry. In addition to the distinct $S(\mathbf{k})$ profiles shown in Figs.~\ref{SK}(c,d) and Figs.~\ref{SkNQS}(c,d), the QSL and KSL phases can be further distinguished by their real-space spin correlations, $F(r)=\langle {{\mathbf{S}} }_{x ,y} \cdot {{\mathbf{S}} }_{x +r,y} \rangle$, as presented in Fig.~\ref{SMSpin}. In the QSL phase [cf. Figs.~\ref{SMSpin}(a-f)], $F(r)$ decays slowly and is well described by a power-law behavior, indicative of gapless excitations and consistent with the proposed Dirac QSL in the $3J$ model. In contrast, $F(r)$ in the KSL phase [cf. Figs.~\ref{SMSpin}(g-i)] decays rapidly and is well fitted by an exponential behavior, with a very short correlation length $\xi_{\mathrm{s}}\sim1$. The distinct behavior observed for the QSL and KSL regimes on the finite YC6 cylinders provides an additional way to distinguish the two phases. Since the present results are restricted to finite-size systems, they do not by themselves determine whether the KSL is gapped or gapless in the two-dimensional limit.

\begin{figure*}[h]
   \includegraphics[width=0.9\textwidth,angle=0]{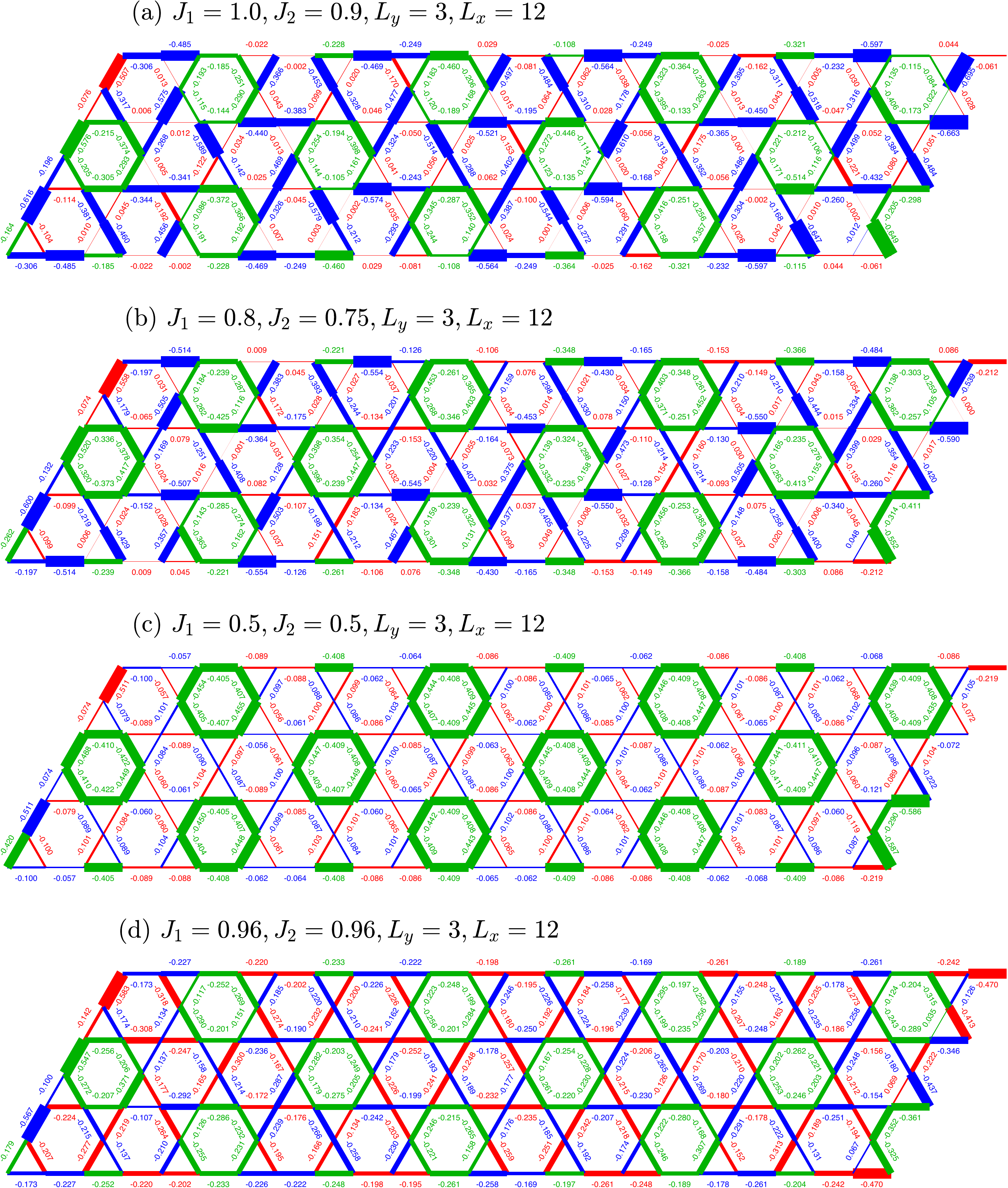}
   \caption{
   \label{SMVBC}
  \textbf{Bond energies in the QSL and KSL phases.} (a-c) Nearest-neighbor bond energies in the QSL phase for the three representative parameter sets, respectively. (d) Nearest-neighbor bond energies in the KSL phase. Different bond colors denote the distinct Heisenberg interactions defined in Eq.~(\ref{eq:model}), while the bond thickness indicates the absolute value of the nearest-neighbor bond energy. All results are obtained on YC6 cylinders with $L_x=12$, using a maximum bond dimension of $D=15000$. 
   }   
\end{figure*}

In Fig.~\ref{SMVBC}, we further examine the possibility of VBC order by calculating the nearest-neighbor bond energies. Figs.~\ref{SMVBC}(a-b) show the spatial distribution of the bond energies on the YC6 kagome lattice in the QSL phase, with model parameters chosen consistently with the finite-temperature simulations. Within the present energy resolution, we find no evidence of broken lattice-translation symmetry, ruling out any discernible VBC order. Similarly, the bond energies in the KSL phase [cf. Fig.~\ref{SMVBC}(d)] are highly uniform, with no discernible signature of VBC order. It is worth noting that along the diagonal line $J_1=J_2$, as $J_1$ and $J_2$ are gradually decrease, the dominant $J_{\hexagon}$ term leads to a characteristic spatial pattern in the bond energies, manifested as the $J_{\hexagon}$-dominated hexagons [green hexagons in Fig.~\ref{SMVBC}(c)]. This behavior is expected, since in the limit of $J_1,J_2\rightarrow0$, the system approaches a collection of decoupled hexagons. Importantly, however, this pattern does not break any symmetry of the $3J$ model Hamiltonian. It merely reflects the modulation of bond energies within the unit cell and therefore does not constitute VBC order.

%%%%%%%%%%%%%%%%%%%%%%%%%%%%%%%%%%%%%%%%%%%%%%%%%%%%%%%%%%%%%%%%%%%%%%%
\section{NQS calculation of the ground state phase diagram}
\label{sec:NQS}

\begin{figure*}[h]
\includegraphics[width=\textwidth]{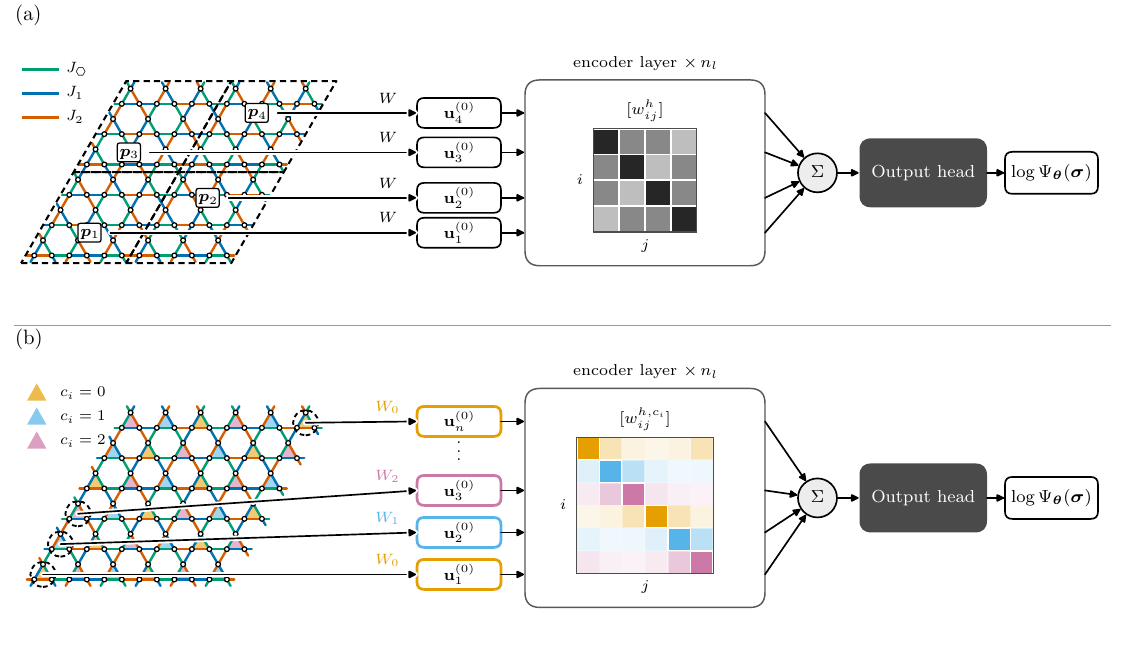}
\caption{\label{fig:arch}
\textbf{Schematic of the two ViT architecture variants.} (a) Architecture used on the $J_1=0.5$ line. The spin configuration on the $6\times6$ torus is divided into $n=4$ patches $\boldsymbol{p}_i$ of $3\times3$ kagome unit cells (shown with dashed outlines), each mapped by the shared linear layer $W$ to an embedding vector $\mathbf{u}^{(0)}_i$ [see Eq.~(\ref{eq:embed})]. The resulting embedding vectors pass through $n_l$ encoder layers whose factored attention weights $w^{h}_{ij}$, Eq.~(\ref{eq:attn}), depend only on the displacement between patches $i$ and $j$ (gray shading). The output vectors are summed over patches and mapped to $\log\Psi_{\boldsymbol{\theta}}(\boldsymbol{\sigma})$ by the output head. (b) Variant used on the $J_1=1$ line. The patches are taken to be single kagome unit cells ($n=36$), colored by their class label $c_i$. The embedding layer $W_{c_i}$ and the attention weights $w^{h,c_i}_{ij}$ are shared only among patches of the same class, Eq.~(\ref{eq:class}).}
\end{figure*}

Our NQS calculations are performed on a torus of $6 \times 6$ kagome unit cells ($N=3\times 6 \times 6=108$ sites) with periodic boundary conditions along both lattice directions, complementing the cylindrical geometries used in our DMRG simulations. To represent the many-body wave function, we use a Vision Transformer (ViT) architecture with factored attention~\cite{Viteritti2025,Viteritti2023,Rende2025_QK,Rende2024_potts}, together with a variant described below. The two are employed in different regions of the phase diagram, according to the translation-symmetry content of the expected phases. We summarize here the main components of the original architecture and refer to Ref.~\cite{Viteritti2025} for a more detailed description. Both variants are illustrated in Fig.~\ref{fig:arch}.

Given an input spin configuration $\boldsymbol{\sigma}=(\sigma^{z}_{1},\ldots,\sigma^{z}_{N})$ in the computational ($S^z$) basis, an NQS outputs the amplitude $\Psi_{\boldsymbol{\theta}}(\boldsymbol{\sigma})=\braket{\boldsymbol{\sigma}|\Psi_{\boldsymbol{\theta}}}\in\mathbb{C}$ of the variational state $\ket{\Psi_{\boldsymbol{\theta}}}$, with $\boldsymbol{\theta}$ denoting the variational parameters. In the ViT of Ref.~\cite{Viteritti2025}, the configuration is divided into $n=N/P$ non-overlapping patches of $P$ spins. Each patch configuration $\boldsymbol{p}_i\in\{\pm1\}^{P}$ is then mapped through a shared linear layer $W$ to a $d$-dimensional embedding vector 
\begin{equation}
\mathbf{u}_i^{(0)} = W(\boldsymbol{p}_i) \equiv E\,\boldsymbol{p}_i + \mathbf{b},
\label{eq:embed}
\end{equation}
where $E\in\mathbb{R}^{d\times P}$ and $\mathbf{b}\in\mathbb{R}^{d}$ are learnable. The resulting sequence of embedding vectors is passed through $n_l$ transformer encoder layers implementing  a factored attention mechanism with $n_h$ heads~\cite{Rende2025_QK,Rende2024_potts}. At each layer $l$, every head $h=1,\dots,n_h$ produces an attention vector 
\begin{equation}
\mathbf{a}^{h}_{i} = \sum_{j=1}^{n} w^{h}_{ij}\,V^{h}(\mathbf{u}_j^{(l)}),
\label{eq:attn}
\end{equation}
for each patch $i$, where $V^h$ is a shared linear layer from $\mathbb{R}^{d}$ to $\mathbb{R}^{d/n_h}$ and  $w^{h}_{ij}$ is the factored attention weight, a learnable scalar depending only on the displacement between patches $i$ and $j$, which renders the attention layer equivariant under patch-level translations of the input. The layer index is suppressed for all quantities other than the embedding vectors.
The $n_h$ attention vectors corresponding to patch $i$ are then concatenated, linearly mixed, and passed through a residual feed-forward step to produce $\mathbf{u}^{(l+1)}_i$~\cite{Viteritti2025}. After the last encoder layer, the resulting vectors $\mathbf{u}_i^{(n_l)}$ are summed over the patch index and mapped to $\log\Psi_{\boldsymbol{\theta}}(\boldsymbol{\sigma})$ by a complex-valued nonlinear layer~\cite{Viteritti2025,Carleo2017}. This final sum-pooling promotes the translation equivariance of the embedding-encoder stack to invariance of the variational state $\ket{\Psi_{\boldsymbol{\theta}}}$ under patch-level translations.  We use this architecture for the scan of the line $J_1=0.5$ in Fig.~\ref{PhaseTransition}(b), choosing a patch of $3\times3$ kagome lattice unit cells ($P=27$), which matches the magnetic unit cell of the $\mathbf{Q}=(1/3,1/3)$ ordered phase expected from the classical phase diagram~\cite{HeringM22} (see the inset of Fig.~\ref{PDDMRG}). This inductive bias is well suited to the magnetically ordered regions, while remaining compatible with disordered states that respect a finer translation symmetry. We use $n_l=4$, $d=160$ and $n_h=40$, corresponding to approximately $1.1\times10^{6}$ parameters.

The second architecture variant is used on the $J_1=1$ line, close to the isotropic kagome limit ($J_1=J_2=J_{\hexagon}$), where the ground state is expected to respect the translation symmetry of the $3J$ Hamiltonian. The patches are fixed to single kagome unit cells ($P=3$), with each patch $i$ being assigned a class label $c_i\in\{0,1,2\}$, recording which of the three inequivalent kagome primitive cells of the nine-site $3J$ unit cell it occupies. The embedding map and the attention weights of Eqs.~(\ref{eq:embed}) and (\ref{eq:attn}) respectively are chosen to be class-dependent,
\begin{equation}
W(\boldsymbol{p}_i) \to  W_{c_i}(\boldsymbol{p}_i), \qquad w^{h}_{ij} \to w^{h,c_i}_{ij},
\label{eq:class}
\end{equation}
replacing their shared counterparts. This change restricts the enforced translation symmetry from all kagome lattice translations, otherwise imposed by single unit cell patches, to only those of the $3J$ Hamiltonian. In addition, given the finer patching, we adopt the spatial attenuation factor of Ref.~\cite{Viteritti2026thermolimit}, which multiplies the attention weights by a normalized exponential of the distance between patches with a learnable decay rate per layer and attention head. This introduces a locality inductive bias, which can be lifted or strengthened through optimization. We use $n_l=8$, $d=72$ and $n_h=12$, corresponding to approximately $4.4\times10^{5}$ parameters.

For every point $(J_1, J_2)$ of the two scans, the ViT parameters $\boldsymbol{\theta}$ are optimized by minimizing the variational energy within the variational Monte Carlo framework~\cite{Becca_2017}. Since the Hamiltonian~(\ref{eq:model}) commutes with $S^z_{\mathrm{tot}}=\sum_i S^z_i$, we perform Monte Carlo sampling within the $S^z_{\mathrm{tot}}=0$ sector using update proposals that exchange pairs of opposite spins. We minimize the energy over $10^4$ steps with $M=8192$ samples per step, using the Subsampled Projected-Increment Natural Gradient Descent (SPRING) optimization algorithm~\cite{Goldshlager2024}, which builds upon the stochastic reconfiguration (SR)~\cite{Sorella1998,Amari1998} and MinSR~\cite{Chen2024,Rende2024} schemes, with momentum $\mu=0.9$ and a cosine-decaying~\cite{Loschilov,hessel2020optax} learning rate from $2 \times 10^{-2}$ to $10^{-3}$. The diagonal shift regularization is held constant at $\lambda=10^{-4}$ for the first variant, while for the second variant it is cosine decayed from $10^{-3}$ to $10^{-4}$ over the run, to stabilize the early optimization of the deeper network. After optimization, the static spin structure factor shown in Figs.~\ref{PhaseTransition} (b,e) and \ref{SkNQS} are estimated with $2^{16}$ Monte Carlo samples drawn from the final state. %All NQS simulations are performed using NetKet~\cite{netket2:2019,netket3:2022}, which is built on top of the JAX~\cite{jax2018github} and Flax~\cite{flax2020github} libraries. %\og{In the NQS calculation, on the $J_1=1$ line, the network enforces the full translation symmetry of the $3J$ Hamiltonian, whereas on the $J_1=0.5$ line it enforces translation symmetry at the level of a $3\times3$ supercell of kagome unit cells, which allows for the $(1/3,1/3)$ magnetic order found in the classical phase diagram~\cite{HeringM22} [cf. Fig.~\ref{fig:arch}].}

%%%%%%%%%%%%%%%%%%%%%%%%%%%%%%%%%%%%%%%%%%%%%%%%%%%%%%%%%%%%%%%%%%%%%%%
\section{Thermal tensor network calculation of the specific heat}
\label{sec:TRG}

\subsection{exponential tensor renormalization group}
The exponential tensor renormalization group (XTRG) method starts by constructing the high‑temperature density operator $\hat{\rho}(\tau) = e^{-\tau H}$, where $\tau$ denotes an infinitesimally small inverse temperature. This initial state can be reliably obtained via the Trotter–Suzuki decomposition~\cite{Trotter1959, Suzuki1990} or through series‑expansion techniques~\cite{Chen2017Series}. A central ingredient of XTRG is its exponential evolution scheme: the density operator is repeatedly squared, i.e., $\hat{\rho}_{n+1} = \hat{\rho}_n \cdot \hat{\rho}_n$, which doubles the inverse temperature at each step. This exponential growth drastically reduces the number of imaginary‑time evolution and truncation steps, thereby enabling accurate simulations down to very low temperatures. In practice, we begin at a tiny inverse temperature $\tau_0$ on the order of  $\sim 10^{-4}$ using the series‑expansion method:

\[
\hat\rho_{0} \equiv \hat\rho(\tau_{0}) \simeq \sum_{k=0}^{\mathcal{N}_{\mathrm{c}}} \frac{(-\tau_{0})^{k}}{k !} H^{k},
\]

with a truncated expansion order $\mathcal{N}_{\mathrm{c}}=4$. We then apply the squaring procedure 19 times to access the lowest temperatures, enabling accurate temperature measurements down to $T\sim 0.05 J_{\hexagon}$.

\begin{figure}[t]
	\centering
    \includegraphics[width=0.8\textwidth,angle=0]{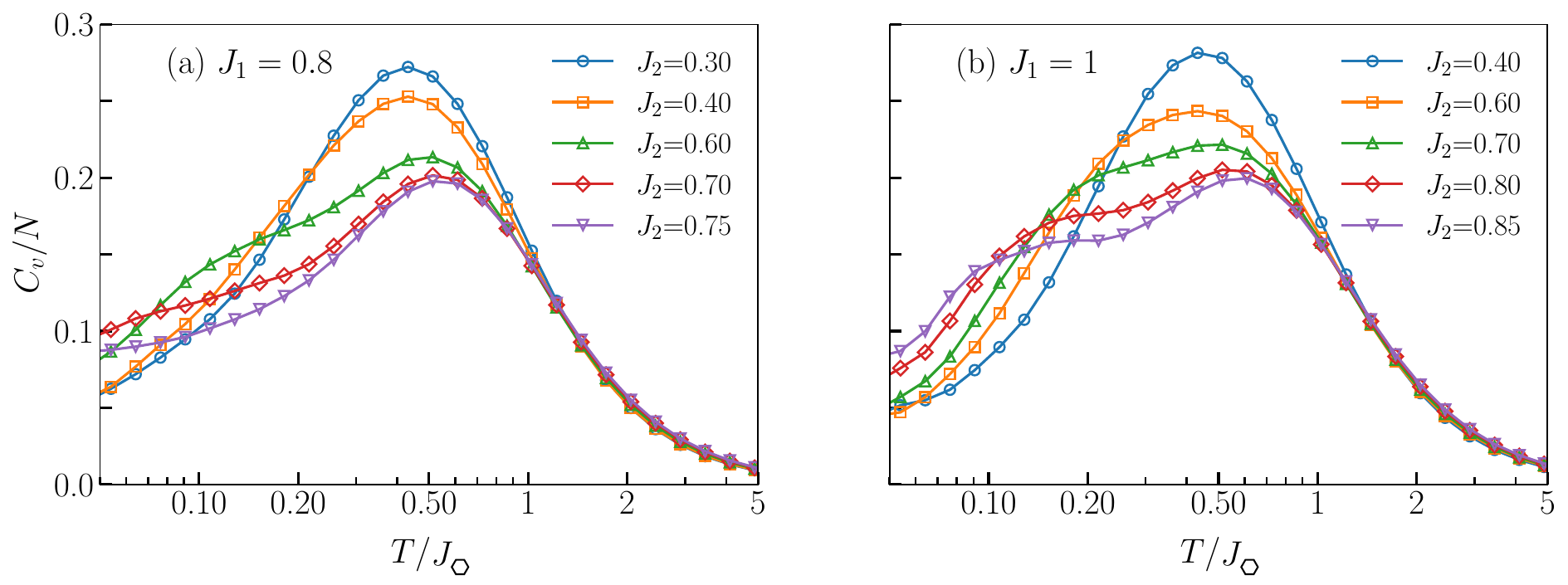}
	\caption{\textbf{Specific heat of the 3$J$ model from XTRG calculation} Specific heat at various $J_2$ values at (a) $J_1=0.8$ and (b) $J_1 = 1$ with bond dimension $D=1000$. The energy scale is set to $J_{\hexagon} = 1$.
	}
	\label{fig:cv_sm}
\end{figure}

In Fig.~\ref{fig:cv_sm}, we present the specific heat at several $J_2$ values computed from XTRG with $D=1000$, for the benefit of interested readers. The energy scale is set to $J_{\hexagon} = 1$. From the $C_v(T)$ curves, distinct features of the ordered and QSL phases are clearly visible for both $J_1 = 0.8$ and $J_1 = 1.0$. The ordered phase is characterized by a pronounced peak near $T \approx 1$, which gradually diminishes as the system enters the QSL phase. Concurrently, a shoulder-like feature emerges at $T \approx 0.1$ within the QSL phase.

\subsection{tangent-space tensor renormalization group}
We also employ the tangent-space tensor renormalization group (tanTRG) method as another self-consistent computation at intermediate temperatures. The simulation procedure also begins by representing the spin Hamiltonian as an MPO, and the thermal density operator $\rho(\beta) = e^{-\beta H}$ is initialized at an infinite-temperature limit (small inverse temperature $\beta = 0$), providing an accurate starting point for the imaginary-time evolution. The cooling process to lower temperatures is then governed by the flow equation $d\rho/d\beta = -H\rho$. To maintain computational efficiency within a manageable bond dimension $D$, the evolution is restricted to the MPO manifold by projecting the right-hand side onto the tangent space at the current state, expressed as $d\rho/d\beta = \mathcal{P}(-H\rho)$, where $\mathcal{P}$ is the tangent-space projector. Throughout the evolution, thermodynamic observables such as the free energy, internal energy, and specific heat are directly evaluated from the MPO representation of the density operator at each temperature step. Practically, we chose a we impose SU(2) symmetry with $D=1000\sim 2000$ U(1) states retained, achieving a peak truncation error of roughly $10^{-4}$ across the temperature range extending to $T\sim 0.08$.

\begin{figure}[t]
	\centering
    \includegraphics[width=0.8\textwidth,angle=0]{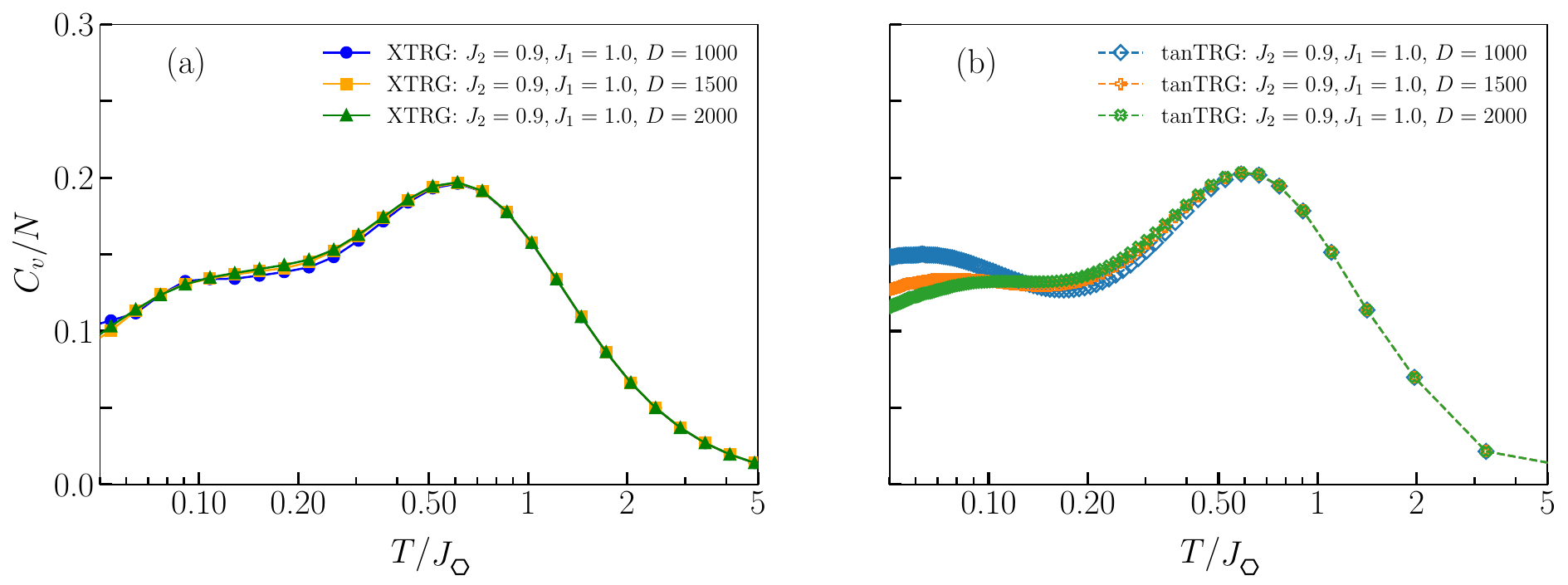}
	\caption{\textbf{Specific heat of the $3J$ kagome models with various bond dimensions.} (a) Specific heat density from XTRG calculations with bond dimensions $D=1000$ (blue), $1500$ (orange), and $2000$ (green). (b) Corresponding tanTRG results with the same bond dimensions (blue, orange, green for $D=1000$, $1500$, and $2000$, respectively). 
	}
	\label{fig:specific_heat_sl}
\end{figure}

Moreover, Fig.~\ref{fig:specific_heat_sl} compares XTRG and tanTRG calculations at $(J_1,J_2)=(1.0,0.9)J_{\hexagon}$ with bond dimensions increasing from $1000$ to $1500$ to $2000$. The XTRG results converge well and show little dependence on bond dimension, whereas the tanTRG results exhibit a noticeable variation. Combined with Fig.~\ref{fig:specific_heat_org}, the tanTRG results gradually approach the XTRG ones, particularly at low temperatures ($T/J_{\hexagon}<0.5$). This behavior indicates that XTRG outperforms tanTRG in the low-temperature regime.

\end{document}